\documentclass[12pt]{article}%
\pdfoutput=1
\usepackage[nosort]{cite}
\usepackage{graphicx}
\usepackage{multicol}
\usepackage{amsfonts}
\usepackage{amssymb}
\usepackage{amsmath}
\usepackage{heck}
\usepackage{setspace}
\usepackage{verbatim}
\usepackage{color}
\usepackage{longtable}
\usepackage{float}
\usepackage{epsfig}
\usepackage{epstopdf}
\usepackage{tikz}
\usepackage[margin=1in]{geometry}
\usepackage{titletoc}%
\usepackage{hyperref}
\hypersetup{colorlinks,%
citecolor=black,%
filecolor=black,%
linkcolor=black,%
urlcolor=black,%
pdftex}
\providecommand{\U}[1]{\protect\rule{.1in}{.1in}}
\usetikzlibrary{decorations.markings}

\numberwithin{equation}{section}

\newcommand{\ba}{\begin{eqnarray}}
\newcommand{\ea}{\end{eqnarray}}

\newcommand{\mf}{\mathfrak}

\newcommand{\ov}{\overset }
\newcommand{\ovo}{\overset{1} }

\newcommand{\be}{\begin{equation}}
\newcommand{\ee}{\end{equation}}

\begin{document}

\date{November 2015}

\title{F-theory and the Classification of Little Strings}

\institution{PERIMETER}{\centerline{${}^{1}$Perimeter Institute for Theoretical Physics, Waterloo, ON N2L 2Y5, CA}}

\institution{HARVARD}{\centerline{${}^{2}$Jefferson Physical Laboratory, Harvard University, Cambridge, MA 02138, USA}}

\institution{UNC}{\centerline{${}^{3}$Department of Physics, University of North Carolina, Chapel Hill, NC 27599, USA}}

\institution{COLUMBIA}{\centerline{${}^{4}$Department of Physics, Columbia University, New York, NY 10027, USA}}

\institution{CUNY}{\centerline{${}^{5}$CUNY Graduate Center, Initiative for the Theoretical Sciences, New York, NY 10016, USA}}

\institution{UCSBmath}{\centerline{${}^{6}$Department of Mathematics, University of California Santa Barbara, CA 93106, USA}}

\institution{UCSBphys}{\centerline{${}^{7}$Department of Physics, University of California Santa Barbara, CA 93106, USA}}

\authors{Lakshya Bhardwaj\worksat{\PERIMETER}\footnote{e-mail: {\tt lbhardwaj@perimeterinstitute.ca}},
Michele Del Zotto\worksat{\HARVARD}\footnote{e-mail: {\tt delzotto@physics.harvard.edu}}, Jonathan J. Heckman\worksat{\UNC, \COLUMBIA, \CUNY}\footnote{e-mail: {\tt jheckman@email.unc.edu}}, \\[2mm]
David R. Morrison\worksat{\UCSBmath, \UCSBphys}\footnote{e-mail: {\tt drm@physics.ucsb.edu}},
Tom Rudelius\worksat{\HARVARD}\footnote{e-mail: {\tt rudelius@physics.harvard.edu}},
and Cumrun Vafa\worksat{\HARVARD}\footnote{e-mail: {\tt vafa@physics.harvard.edu}}}

\abstract{Little string theories (LSTs) are UV complete non-local 6D theories decoupled
from gravity in which there is an intrinsic string scale. In this paper we present
a systematic approach to the construction of supersymmetric LSTs
via the geometric phases of F-theory. Our central result is that all LSTs with more than one tensor multiplet are obtained by a mild extension of 6D superconformal field theories (SCFTs) in which the theory is supplemented by an additional, non-dynamical tensor multiplet, analogous to adding an affine node to an ADE quiver, resulting in a negative semidefinite Dirac pairing.
We also show that all 6D SCFTs naturally embed in an LST. Motivated by physical considerations, we
show that in geometries where we can verify the presence of two elliptic fibrations, exchanging the roles
of these fibrations amounts to T-duality in the 6D theory compactified on a circle.}

\maketitle

\tableofcontents

\enlargethispage{\baselineskip}

\setcounter{tocdepth}{2}

\newpage

\section{Introduction \label{sec:INTRO}}

One of the concrete outcomes from the post-duality era of string theory is the wealth of insights it
provides into strongly coupled quantum systems. In the context of string compactification,
this has been used to argue, for example, for the existence of
novel interacting conformal field theories in spacetime dimensions $D > 4$. In
a suitable gravity-decoupling limit, the non-local ingredients of a theory of extended
objects such as strings are instead captured by a quantum field theory with a local stress energy tensor.

String theory also predicts the existence of novel non-local theories. Our focus in this work will be on
6D theories known as little string theories (LSTs).\footnote{We leave open
the question of whether non-supersymmetric little string theories exist.} For a partial list of
LST constructions, see e.g.  \cite{Witten:1995zh,Aspinwall:1996vc,Aspinwall:1997ye,Seiberg:1997zk,Intriligator:1997dh, Hanany:1997gh, Brunner:1997gf}. In these systems, 6D gravity is decoupled, but an intrinsic string scale $M_{string}$ remains.
At energies far below $M_{string}$, we have an effective theory which is well-approximated by
the standard rules of quantum field theory with a high scale cutoff. However,
this local characterization breaks down as we reach the scale $M_{string}$. The UV completion, however, is not a
quantum field theory.\footnote{In fact, all
known properties of LSTs are compatible
with the axioms for quasilocal quantum field theories \cite{Kapustin:1999ci}.}

The mere existence of little string theories leads to a tractable setting for studying many of the essential features of string
theory --such as the presence of extended objects-- but with fewer complications (such as coupling to quantum gravity). It also raises
important conceptual questions connected with the UV completion of low energy quantum field theory. For
example, in known constructions these theories exhibit T-duality upon toroidal compactification \cite{Seiberg:1997zk,Intriligator:1997dh,Intriligator:1999cn} and a Hagedorn density of states \cite{Aharony:1998ub},
properties which are typical of closed string theories with tension set by $M_{string}^2$.

Several families of LSTs have been engineered in the context of superstring theory by using various combinations of branes probing geometric singularities. The main idea in many of these constructions is to take a gravity-decoupling
limit where the 6D Planck scale $M_{pl} \rightarrow \infty$ and the string coupling $g_{s} \rightarrow 0$, but with an effective string scale
$M_{string}$ held fixed. Even so, an overarching picture of how to construct (and study) LSTs has remained somewhat elusive.

Our aim in this work is to give a systematic approach for realizing
LSTs via F-theory and to explore its consequences.\footnote{More precisely, we focus on the case of geometric phases of F-theory, ignoring the (small) list of possible models with ``frozen'' singularities (see e.g. \cite{Witten:1997bs, deBoer:2001px, Tachikawa:2015wka}). We return to this point later in section \ref{sec:OUTLIERS} when we discuss the mismatches between field theory motivated LST constructions and their possible lifts to string constructions.} To do this, we will use both a bottom up characterization of little string theories on the tensor branch (i.e. where all effective strings have picked up a tension), as well as a formulation in terms of compactifications of F-theory. To demonstrate UV completeness of the resulting models we will indeed need to use the F-theory characterization.

Recall that in F-theory, we have a non-compact base $B$ of complex dimension two, which is supplemented by an elliptic fibration to reach a non-compact Calabi-Yau threefold. In the resolved phase, the intersection pairing of the base coincides with the Dirac pairing for two-form potentials of the theory on its tensor branch. For an SCFT, we demand that the Dirac pairing is negative definite. For an LST, we instead require that this pairing is negative semidefinite, i.e., we allow for a non-trivial null space.

F-theory also imposes the condition that we can supplement this base by an appropriate elliptic fibration to reach a non-compact Calabi-Yau threefold. In field theory terms, this is usually enforced by the condition that all gauge theoretic anomalies are cancelled on the tensor branch of the theory. For 6D gauge theories which complete to LSTs, this condition was discussed in detail in reference \cite{Seiberg:1996qx}. Even when no gauge theory interpretation is available, this means that in the theory on the tensor branch, some linear combination of tensor multiplets is non-dynamical, and instead defines a dimensionful parameter (effectively a UV cutoff) for the 6D effective field theory.

\begin{figure}[t!]
    \centering
        \includegraphics[scale=.7]{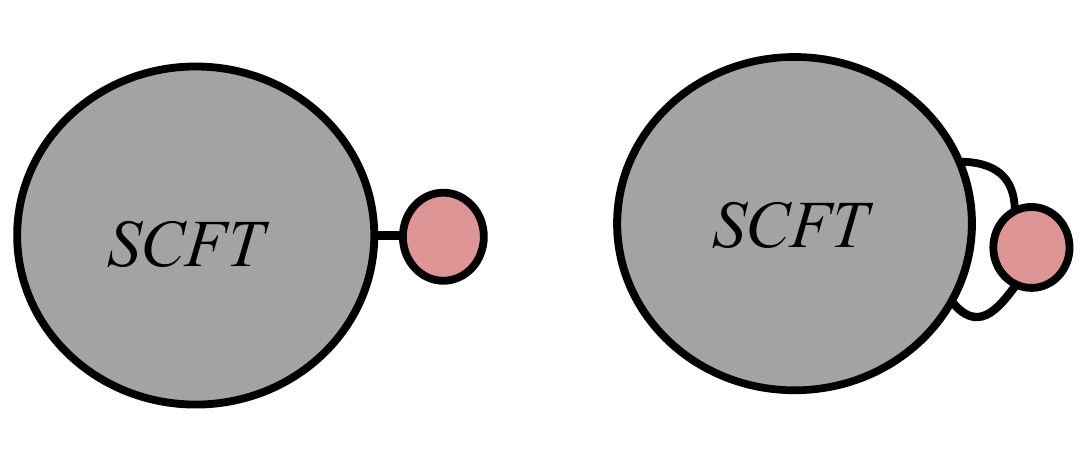}
            \caption{Depiction of how to construct the base of an F-theory model for an LST. All LST bases are obtained by adding one additional
            curve to the base for a 6D SCFT. This additional curve can intersect either one or two curves of the SCFT base. Much as in the study of Lie algebras, LSTs should be viewed as an ``affine extension'' of SCFTs.}
    \label{fig:affinizer}
\end{figure}

In F-theory terms, classifying LSTs thus amounts to determining all possible elliptic Calabi-Yau threefolds which support a base $B$ with negative semidefinite intersection pairing. One of our results is that all LSTs are given by a small extension of 6D SCFTs, i.e. they can always be obtained
by adding just one more curve to the base of an SCFT so that the resulting base has an intersection pairing with a null direction. Put in field theory terms, we find that the string charge lattice of any LST with more than one tensor multiplet is an affine extension of the string charge lattice of an SCFT, with the minimal imaginary root of the lattice corresponding to the little string charge. Hence, much as in the case of Lie algebras, all LSTs arise from an affine extension of SCFTs. See figure \ref{fig:affinizer} for a depiction of this process.

In fact, the related classification of 6D SCFTs has already been successfully carried out. See e.g. the partial list of references \cite{Heckman:2013pva, Gaiotto:2014lca, DelZotto:2014hpa, Heckman:2014qba, DelZotto:2014fia, Heckman:2015bfa, Bhardwaj:2015xxa}.
What this means is that we can freely borrow this structure to establish a classification of LSTs.
Much as in reference \cite{Heckman:2015bfa}, we establish a similar ``atomic classification'' of how LSTs are built up from smaller constitutent elements. We find that the base of an F-theory geometry is organized according to a single spine of ``nodes'' which are decorated by possible radicals, i.e. links which attach to these nodes. As opposed to the case of SCFTs, however, the topology of an LST can be either a tree or a loop.

Using this characterization of LSTs, we also show that all 6D SCFTs can be embedded in some LST
by including additional curves and seven-branes:
\begin{equation}
\mathrm{6D\, SCFTs} \rightarrow \mathrm{6D \, LSTs}.
\end{equation}
Deformations in both K\"ahler and complex structure moduli for the LST then take us back to the original SCFT. It is curious to note that although many 6D SCFTs cannot be coupled to 6D supergravity, they \textit{can} always be embedded in another theory with an intrinsic length scale.

A hallmark of all known LSTs is T-duality, that is, by compactifying on a small circle,\footnote{That is, small when compared with the effective string scale.} we reach another 6D LST compactified on a circle of large radius. This motivates a physical conjecture that all LSTs exhibit such a T-duality. In geometries where we can verify the presence of two elliptic fibrations, we find that exchanging the roles
of these fibrations amounts to T-duality in the 6D theory compactified on a circle.\footnote{This has been independently observed by Daniel Park \cite{ParkConversation}.}
In some cases, we find that T-duality takes us to the \textit{same} LST. For  a recent application of this double elliptic fibration structure
in the study of the correspondence between instantons and monopoles via compactifications of little string theory, see reference \cite{Hohenegger:2015btj}.

The rest of this paper is organized as follows. In section \ref{sec:BOTTOM} we state necessary bottom up conditions to realize a LST. This includes the core condition that the Dirac pairing for an LST is a negative semidefinite matrix. After establishing some of the conditions this enforces, we then turn in section \ref{sec:LSTF} to the rules for constructing LSTs in F-theory. We also explain the (small) differences between the rules for constructing LSTs versus SCFTs. Section \ref{sec:EXAMPLES} gives some examples of known constructions of LSTs, and their embedding in F-theory. In section \ref{sec:DECOUP} we show how decoupling a tensor multiplet to reach an SCFT leads to strong constraints on possible F-theory models. In section \ref{sec:ATOMIC} we present an atomic classification of bases, and in section \ref{sec:FIBERS} we turn to the classification of possible
elliptic fibrations over a given base. In section \ref{sec:EMBED} we demonstrate that every 6D SCFT constructed in F-theory can be embedded into at least one 6D LST constructed in F-theory. In section \ref{sec:Tduality} we show how T-duality of the LST shows up as the existence of a double elliptic fibration structure, and the exchange in the roles of the elliptic fibers.
As a consequence, we show that LSTs can acquire discrete gauge symmetries
for particular values of their moduli.
In section \ref{sec:OUTLIERS} we discuss the small mismatch with possible LST constructions suggested by field theory, and their potential embedding in a non-geometric phase of an F-theory model.
Section \ref{sec:CONC} contains our conclusions, and some additional technical material is deferred to a set of Appendices.

\section{LSTs from the Bottom Up \label{sec:BOTTOM}}

In this section we state some of the conditions necessary to realize a supersymmetric little string theory.

We consider 6D supersymmetric theories which admit a tensor branch (which can be zero dimensional, as will be the case for many LSTs), that is, we will have a theory with some dynamical tensor multiplets, and vacua parameterized (at low energies) by vevs of scalars in these tensor multiplets. We will tune the vevs of the dynamical scalars to zero to reach a point of strong coupling. Our aim will be to seek out theories in which this region of strong coupling is not described by an SCFT, but rather, by an LST. In addition to dynamical tensor multiplets, we will allow the possibility of non-dynamical tensor multiplets which set mass scales for the 6D supersymmetric theory.

Recall that in a theory with $T$ tensor multiplets, we have scalars $S^{I}$ and their bosonic superpartners $B^{-,I}_{\mu \nu}$, with anti-self-dual field strengths. The vevs of the $S^{I}$ govern, for example, the tension of the effective strings which couple to these two-form potentials. In a theory with gravity, one must also include an additional two-form potential $B_{\mu \nu}^{+}$ coming from the graviton multiplet. Given this collection of two-form potentials, we get a lattice of string charges $\Lambda_{string}$, and a Dirac pairing:\footnote{Here we ignore possible torsional contributions to the pairing.}
\begin{equation}
\Lambda_{string} \times \Lambda_{string} \rightarrow \mathbb{Z},
\end{equation}
in which we allow for the possibility that there may be a mull space for this pairing.  It is convenient to describe the pairing in terms of a matrix $A$ in which all signs have been reversed.  Thus, we can
write the signature of $A$ as $(p,q,r)$ for $q$ self-dual field strengths, $p$ anti-self-dual field strengths, and $r$ the dimension of the null space.

Now, in a 6D theory with $q$ self-dual field strengths and $p$ anti-self-dual field strengths, the signature of $A$ is $(p,q,0)$. For a 6D supergravity theory with $T$ tensor multiplets, the signature is $(T,1,0)$. In fact, even more is true in a 6D theory of gravity: diffeomorphism invariance enforces the condition found in \cite{Seiberg:2011dr} that $\det A = -1$.

Now, since we are interested in supersymmetric theories decoupled from gravity we arrive at the \textit{necessary} condition that the signature of $A$ is $(p,0,r)$. In this special case, each of our two-form potentials has a real scalar superpartner, which we denote as $S^{I}$. The kinetic term for these scalars is:
\begin{equation}
\mathcal{L}_{eff} \supset A_{IJ} \partial S^{I} \partial S^{J}.
\end{equation}
Observe that if $A$ has a zero eigenvector, some linear combinations of the scalars will have a trivial kinetic term.
When this occurs, these tensor multiplets define parameters of the effective theory
on the tensor branch (i.e. they are non-dynamical fields).

This leaves us with two general possibilities. Either $A$ is positive definite (i.e. $A > 0$), or it is positive semidefinite (i.e. $A \geq 0$). Recall, however, that to reach a 6D SCFT, a necessary condition is $A > 0$ \cite{Seiberg:1996qx, Heckman:2013pva, Heckman:2015bfa, Bhardwaj:2015xxa}.
We summarize the various possibilities for self-consistent 6D theories:
\begin{equation}%
\begin{tabular}
[c]{|c|c|c|c|}\hline
& 6D SUGRA & 6D LST & 6D SCFT\\\hline
signature & $(T,1,0)$ & $(p,0,r)$ & $(T,0,0)$\\\hline
$\det A:$ & $ \det A  = - 1$ & $\det A=0$ & $\det
A>0$\\\hline
\end{tabular}
\end{equation}
For now, we have simply indicated an LST as any theory where $\det A = 0$.

As already mentioned, when $\det A = 0$, some linear combinations of the scalar fields for tensor multiplets will have trivial kinetic term.
This means that they are better viewed as defining dimensionful
parameters. For example, in the case of a 6D theory with a single gauge group factor and
no dynamical tensor multiplets, this parameter is just the overall value $S_{null} = 1 / g_{YM}^2$, with $g_{YM}$ the Yang-Mills coupling of a gauge theory. Indeed, this Yang-Mills theory contains solitonic solutions which we can identify with strings:
\begin{equation}\label{soliton}
F = - \ast_{4} F,
\end{equation}
that is, we dualize in the four directions transverse to an effective string.
More generally, we can expect $A$ to contain some general null space, and
with each null direction, a non-dynamical tensor multiplet of parameters:
\begin{equation}
\overrightarrow{v}_{null} \equiv N_{1} \overrightarrow{v}^{1} + ... + N_{T} \overrightarrow{v}^{T} \,\,\,\text{such that} \,\,\, A \cdot \overrightarrow{v}_{null} = 0.
\end{equation}
for the two-form potential, and:
\begin{equation}
S_{null} = N_{1} S^{1} + ... + N_{T}S^{T}
\end{equation}
for the corresponding linear combination of scalars. Since they specify
dimensionful parameters, we get an associated mass scale, which
we refer to as $M_{string}$:
\begin{equation}
S_{null} = M_{string}^2.
\end{equation}
Returning to our example from 6D gauge theory, the tension of the solitonic string
in equation (\ref{soliton}) is just $1/g_{YM}^2 = M_{string}^2$. At energies
above $M_{string}$, our effective field theory is no longer valid, and we must provide a
UV completion.

On general grounds, $A \geq 0$ could have many null directions. However, in the case where where we have a single interacting theory, i.e. when $A$ is simple, there are further strong restrictions. As explained in reference \cite{MR0302779}, when $A \geq 0$ is simple, all of its minors are positive definite: $A_{minor} > 0$. Consequently, there is precisely \textit{one} zero eigenvalue, and the eigenvector is a positive linear combination of basis vectors. Consequently, there is only one dimensionful parameter $M_{string}$. This also means that if we delete any tensor multiplet, we reach a positive definite intersection pairing, and consequently, a 6D SCFT. What we have just learned is that if we work in the subspace orthogonal to the ray swept out by $S_{null}$, then the remaining scalars can all be collapsed to the origin of moduli space. When we do this, we reach the LST limit.

We shall refer to this property of the matrix $A$ as the ``tensor-decoupling criterion'' for an LST. As we show in subsequent sections, the fact that decoupling any tensor multiplet takes us to an SCFT imposes sharp restrictions.

Even so, our discussion has up to now focussed on some necessary conditions to reach a UV complete theory different from a 6D SCFT.
In references \cite{Seiberg:1996qx, Bhardwaj:2015xxa} the specific case of 6D supersymmetric gauge theories was considered, and closely related consistency conditions for UV completing to an LST were presented. Here, we see the same consistency condition $A \geq 0$ appearing for \textit{any} effective theory with (possibly non-dynamical) tensor multiplets.

Indeed, simply specifying the tensor multiplet content provides an incomplete characterization of the tensor branch. In addition to this, we will also have vector multiplets and hypermultiplets. For theories with only eight real supercharges, anomaly cancellation often imposes tight consistency conditions.

There is, however, an important difference in the way anomaly cancellation operates in a 6D SCFT compared with a 6D LST. The crucial point is that because $A$ has a zero eigenvalue, there is a non-dynamical tensor multiplet which does not participate in the Green-Schwarz mechanism. In other words, on the tensor branch of an LST with $T$ tensor multiplets, at most only $T-1$ participate. This is not particularly worrisome since as explained in reference \cite{Seiberg:2011dr} and further explored in reference \cite{DelZotto:2015isa}, there is in general a difference between the tensor multiplets which participate in anomaly cancellation and those which appear in the tensor branch of a general 6D theory.

Though we have given a number of \textit{necessary} conditions that any putative LST must satisfy, to truly demonstrate their existence we must pass beyond effective field theory, embedding these theories in a UV complete framework such as string theory. We therefore now turn to the F-theory realization of little string theories.

\section{LSTs from F-theory \label{sec:LSTF}}

In this section we spell out the geometric conditions necessary to realize
LSTs in\ F-theory. Recall that in a little string theory, we
are dealing with a 6D theory which contains strings with finite
tension. As such, they are an intermediate case between the case of a 6D
superconformal field theory (which only contains tensionless strings), and the
full string theory (i.e., one in which gravity is dynamical).

Any supersymmetric F-theory compactification to six dimensions is defined by an elliptically fibered Calabi-Yau threefold $X\rightarrow B$. Here, $X$ is the total space and $B$ is the base. The elliptic fibration can be described by a local Weierstrass model
\begin{equation}
y^{2}=x^{3}+f x+g,
\end{equation}
where $f$ and $g$ are local functions on $B$, that globally are sections respectively of $\mathcal{O}_{B}(-4K_{B})$ and $\mathcal{O}_{B}(-6K_{B})$, $K_B$ being the canonical class of $B$. The discriminant of the elliptic fibration is:
\begin{equation}
\Delta \equiv 4f^{3} + 27g^{2}
\end{equation}
which globally is a section of $\mathcal{O}_{B}(-12K_B)$. The discriminant locus $\Delta=0$ is a divisor, and its irreducible components tell us the locations of degenerations of elliptic fibers. Such singularities determine monodromies for the complex structure parameter $\tau$ of the elliptic fiber, which is interpreted in type IIB string theory as the axio-dilaton field. In type IIB language, the discriminant locus signals the location of seven-branes in the F-theory model.

In F-theory, decoupling gravity means we will always be dealing with a non-compact base $B$. When
all curves of $B$ are of finite non-zero size, we get a 6D effective theory with a lattice of string charges:
\begin{equation}
\Lambda_{string} = H_{2}^{cpct}(B,\mathbb{Z}).
\end{equation}
The intersection form defines a canonical pairing:
\begin{equation}
A_{intersect} : \Lambda_{string} \times \Lambda_{string} \rightarrow \mathbb{Z},
\end{equation}
which we identify with the Dirac pairing:
\begin{equation}
A_{Dirac} = A_{intersect}.
\end{equation}
We also introduce the ``adjacency matrix''
\begin{equation}
A_{adjacency} = - A_{Dirac}.
\end{equation}
To streamline the notation, we shall simply denote the adjacency matrix as $A$. The two-form potentials of the 6D theory arise from reduction of the four-form potential of type IIB string theory. Additionally, the volumes of the various compact two-cycles translate to the real scalars of tensor multiplets:
\begin{equation}
S^{I} \propto \mathrm{Vol}(\Sigma_{I}).
\end{equation}
In the F-theory model, the appearance of a null vector for $A_{intersect}$
means that some of these moduli are not dynamical in the 6D effective field theory.
Rather, they define dimensionful parameters / mass scales. This follows from the Grauert-Artin contractibility criterion in algebraic geometry \cite{Grauert, Artin}, which states that any given curve in a complex surface is contractible if and only if the intersection matrix of its irreducible components is negative definite. This simple geometrical criterion gives a necessary condition ($A>0$) for engineering SCFTs and implies that any null eigenvalues of $A$ correspond to non-contractible curves, which thus define intrinsic energy scales.

To define an F-theory model, we need to ensure that there is an elliptic Calabi-Yau $X$ in which $B$ is the base. A necessary condition for realizing the existence of an elliptic model is that the collection of curves entering in a base $B$ are obtained by gluing together the ``non-Higgsable clusters'' (NHCs) of reference \cite{Morrison:2012np}
via $\mathbb{P}^1$'s of self-intersection $-1$.

Recall that the non-Higgsable clusters are given by collections of up to three
$\mathbb{P}^{1}$'s in which the minimal singular fiber type is dictated by the
self-intersection number of the $\mathbb{P}^{1}$. The self-intersection
number, and associated gauge symmetry and matter content are as follows:%
\begin{align}
&
\begin{tabular}
[c]{|l|l|l|l|l|l|l|}\hline
Self-intersection & $-3$ & $-4$ & $-5$ & $-6$ & $-7$ & $-8$\\\hline
Gauge Theory & $\mathfrak{su}_{3}$ & $\mathfrak{so}_{8}$ & $\mathfrak{f}_{4}$
& $\mathfrak{e}_{6}$ & $\mathfrak{e}_{7}\oplus$ $\frac{1}{2}56$ &
$\mathfrak{e}_{7}$\\\hline
\end{tabular}
\\
&
\begin{tabular}
[c]{|l|l|l|l|l|}\hline
Self-intersection & $-9$ & $-10$ & $-11$ & $-12$\\\hline
Gauge Theory & $\mathfrak{e}_{8}\oplus$ 3 inst & $\mathfrak{e}_{8}\oplus$ 2
inst & $\mathfrak{e}_{8}\oplus$ 1 inst & $\mathfrak{e}_{8}$\\\hline
\end{tabular}
\\
&
\begin{tabular}
[c]{|l|l|l|l|}\hline
Self-intersection & $-3,-2$ & $-2,-3,-2$ & $-3,-2,-2$\\\hline
Gauge Theory & $%
\genfrac{}{}{0pt}{}{\mathfrak{g}_{2}\times\mathfrak{su}_{2}}{\oplus\frac{1}%
{2}(7+1,2)}%
$ & $%
\genfrac{}{}{0pt}{}{\mathfrak{su}_{2}\times\mathfrak{so}_{7}\times
\mathfrak{su}_{2}}{\oplus\frac{1}{2}(2,8,1)\oplus\frac{1}{2}(1,8,2)}%
$ & $%
\genfrac{}{}{0pt}{}{\mathfrak{g}_{2}\times\mathfrak{sp}_{1}}{\oplus\frac{1}%
{2}(7,2)\oplus\frac{1}{2}(1,2)}%
$\\\hline
\end{tabular}
\end{align}
in addition, we can also consider a single $-1$ curve, and configurations of
$-2$ curves arranged either in an ADE\ Dynkin diagram, or its affine extension
(in the case of little string theories). The local rules for building up an
F-theory base compatible with these NHCs amount to a local gauging condition
on the flavor symmetries of a $-1$ curve: We scan over product subalgebras of
the $\mathfrak{e}_{8}$ flavor symmetry which are also represented by the
minimal fiber types of the NHCs. When they exist, we get to \textquotedblleft
glue\textquotedblright\ these NHCs together via a $-1$ curve.

For a general elliptic Calabi-Yau threefold, the curves appearing in a given gluing configuration
can lead to rather intricate intersection patterns. For example, two curves may intersect more than once, and may therefore form
either a closed loop, or an intersection with some tangency. Additionally, we may have three curves all meeting at the same point, as in the
case of the type $IV$ Kodaira fiber. Finally, a single $-1$ curve may in general
intersect more then just two curves. The possible ways to locally glue together such NHCs has also been worked out explicitly in
reference \cite{Morrison:2012np} (see also \cite{Morrison:2012js}). The main idea, however, is that since the $-1$ curve theory defines a
6D SCFT with $E_8$ flavor symmetry, we must perform a gluing compatible
with gauging some product subalgebra of the Lie algebra $\mathfrak{e}_8$.

What this means in general is that the adjacency matrix provides only a partial characterization of intersecting curves in the base of a geometry.
To handle these different possibilities, we therefore introduce the following notation:
\begin{align}
\text{Normal Intersection}  & \text{: }a,b\,\,\, \text{or}\,\,\, ab \\
\text{Tangent Intersection}  & \text{: }a||b\\
\text{Triple Intersection}  & \text{: }a\overset{b}{\bigtriangledown}c\\
\text{Looplike configuration}  & \text{: }//a_{1}...a_{k}//.
\end{align}

Now, decoupling gravity to reach an SCFT or an LST leads to significant restrictions on the possible ways
to glue together NHCs. In the case of a 6D SCFT, contractibility of all curves in the base means first, that all of the compact curves are $\mathbb{P}^{1}$'s, and further, that a $-1$ curve can intersect at most two other curves. Additionally,
all off-diagonal entries of the intersection pairing are either zero or one. In the case of LSTs, however,
the curves of the base could include a $T^2$, and a $-1$ curve can potentially intersect more than two curves.
Additionally, there is also the possibility that the off-diagonal entries of the adjacency matrix may be different than just zero or one.

Again, we stress that the intersection pairing provides only \textit{partial} information.
For example, a curve of self-intersection zero could refer either to a $\mathbb{P}^1$, or to a $T^2$.
In the case of a $T^2$ of self-intersection zero, the normal bundle need not be trivial, but could be
a torsion line bundle instead.
Additionally, an off-diagonal entry in the adjacency matrix which is two may either refer to a pair of curves which intersect twice, or to a single intersection of higher tangency. The case of tangent intersections violates the condition of normal crossing (which is known to hold for SCFTs \cite{Heckman:2013pva}
but fails for LSTs). An additional type of normal crossing violation  appears when we blow down a $-1$ curve meeting more than two curves. In Appendix \ref{Appendix:Singularity} we determine the types of matter localized when there are violations of normal crossing.

\subsection{Geometry of the Gravity-Decoupling Limit}

We now discuss how to obtain limits of F-theory compactifications in which
gravity is decoupled, following a program initiated in
\cite{Beasley:2008kw},
worked out in detail in
\cite{Cordova:2009fg} (see also \cite{Heckman:2010pv}),
and extended to the case of 6D SCFTs in
\cite{global-symmetries}.
For this purpose, we consider F-theory from the perspective of the type
IIB string, with the volume of the base $B$ of the F-theory compactification
providing a Planck scale for the compactified theory.
We will see that the quest for decoupled gravity leads to the same
condition on semidefiniteness of the intersection matrix of the compact
curves, and moreover we will see how to ensure that the F-theory base
$B$ in such cases has a metric of the appropriate kind.

\subsubsection{The Case of Compact Base}

We begin with the
case in which the F-theory base $B$ is a compact surface, and suppose we have
a sequence of metrics (specified by their K\"ahler forms $\omega_i$)
which decouple gravity in the limit $i\to \infty$.  In particular,
the volume must go to infinity:  $\lim_{i\to\infty}\vol(\omega_i)=\infty$.

To investigate the geometry of this family of metrics, we temporarily
rescale them and consider the K\"ahler forms
\begin{equation}
\widetilde{\omega}_i := \frac{\omega_i}{\sqrt{\vol(\omega_i)}}.
\end{equation}
The rescaled metrics all have volume $1$, and since the closure of the set of
volume $1$ K\"ahler classes on $B$ is compact, there must be a convergenct
subsequence of K\"ahler classes $[\widetilde{\omega}_{i_j}]$ whose limit
\begin{equation}
[\widetilde{\omega}_\infty] = \lim_{j\to\infty} [\widetilde{\omega}_{i_j}]
\label{eq:rescaled-limit}
\end{equation}
lies in the closure of the K\"ahler cone.  If the original
sequence was chosen generically, the limit of the rescaled sequence
will be an interior point of the K\"ahler cone, and in this case
all areas and volumes grow uniformly as we take the limit of the
original sequence $\omega_i$.  Gravity decouples, but all other
physical quantities measured by areas and volumes approach either
zero or infinity, leaving us with a trivial theory.

However, if the rescaled limit \eqref{eq:rescaled-limit} lies on the
boundary of the K\"ahler cone, more interesting things can happen.
In favorable
circumstances, such as those present in Mori's cone theorem \cite{MR662120}
and its generalizations \cite{kawamata-cone}, we can form another complex
space $\overline{B}$ out of $B$ by identifying pairs of points $p$ and $q$ whenever
they are both contained in a curve $C$ whose area vanishes in
the limit.  There is a holomorphic map $\pi:B\to \overline{B}$ for which all such
curves of zero limiting area are contained in fibers $\pi^{-1}(t)$,
$t\in\overline{B}$.

As already pointed out in
\cite{Cordova:2009fg}, there are two qualitatively different cases:
$\overline{B}$ might be a  surface or it might be a curve.  (It is not
possible for $\overline{B}$ to be a point since there are some curves
$C\subset B$ whose area does not vanish in the limit.)
If $\overline{B}$ is a surface, then the map $\pi:B \to
\overline{B}$ contracts some curves to points,
and may create singularities in
$\overline{B}$.  It is widely
believed, and has been mathematically
proven under certain hypotheses \cite{MR3261011,2015arXiv150705082D}, that the
limiting metric $\widetilde{\omega}_\infty$ can be interpreted as a metric
$\omega_{\overline{B}}$
on the smooth part of $\overline{B}$.

On the other hand, if $\overline{B}$ is a curve, so that $\pi$ has
curves $\Sigma_t$, $t\in\overline{B}$ as fibers, then we again
 expect the limiting metric $\widetilde{\omega}_\infty$
to be induced by a metric on $\overline{B}$, although there are fewer mathematical
theorems covering this case.  (See \cite{GrossWilson:LCSL} for one
known theorem of this kind.)

In general, we do not expect the curves contracted by $\pi$ to
necessarily have zero area in the gravity-decoupling limit.  This
can be achieved by starting with a reference K\"ahler form $\omega_0$
on $B$ as well as a (possibly degenerate) K\"ahler form
$\omega_{\overline{B}}$ on $\overline{B}$, and constructing a limit of the form
\begin{equation}
\lim_{t\to\infty}\left( \omega_0+t\pi^*(\omega_{\overline{B}})\right).
\end{equation}
In the case of an SCFT, we wish all curves contracted by $\pi$ to be
at zero area in the limit, so in that case
we should omit $\omega_0$ and simply
scale up $\pi^*(\omega_{\overline{B}}))$.

\subsubsection{The Case of Non-Compact Base}

Our discussion of the compact bases makes it clear that the decoupling
limit only depends on the metric in a (finite volume)
neighborhood of a collection of
curves on the original F-theory base $B$, together with a rescaling
which takes that neighborhood to infinite volume and smooths out its features
in the process.  This analysis can be applied to an arbitrary base,
compact or non-compact.

If the collection of curves is disconnected, the corresponding points
on $\overline{B}$  to which the collection is mapped will
be moved infinitely far apart during the rescaling process, thus leading
to several decoupled quantum theories.  So to study a single theory, it
suffices to consider a connected configuration.  To reiterate the two
cases we have found:
\begin{enumerate}
\item We may have a connected collection of curves $\Sigma_j$ which can be
simultaneously contracted to a singular point on a space $\overline{B}$.
(The contractibility implies that the intersection matrix is negative
definite.)
When the metric on $\overline{B}$ is rescaled, gravity is decoupled giving a 6D
SCFT (in which every curve in the collection remains at zero area).

Alternatively, we can combine this rescaled metric with another
reference metric which provides finite area to each $\Sigma_j$.  This
produces a quantum field theory in the Coulomb branch of the SCFT.
\item
Or we may have a connected collection of curves $\Sigma_j$ which are
all contained in a single fiber of a map $\pi:B\to \overline{B}$, and include
all components of that fiber.  (This implies that the intersection matrix
is negative semidefinite, with a one-dimensional zero eigenspace.)  We combine
a reference metric on $B$ that sets the areas of the individual $\Sigma_j$'s
with a metric on $\overline{B}$ which is rescaled to decouple gravity,
yielding a little string theory (with the string provided by a D3-brane
wrapping the entire fiber).
The overall area of the fibers of $\pi$ sets the string scale, and the
possible areas of the $\Sigma_j$'s map out the moduli space of the theory.
Gravity is decoupled, and we find an LST.
\end{enumerate}
Note that in the second case, there are two distinct possibilities for
the fibers of the map $\pi$:  the general fiber can be a curve of genus
$0$ or a curve of genus $1$.  In the case of genus $1$, it is possible
for the central fiber to have a nontrivial multiplicity, that is, the
fiber can take the form $m\Sigma$ for some $m>1$.

\section{Examples of LSTs \label{sec:EXAMPLES}}

In the previous section we gave the general rules for constructing LSTs.
Our plan in this section will be to show how the F-theory realization allows us to recover well-known examples
of LSTs previously encountered in the literature.

To this end, we begin by first showing how LSTs with sixteen supercharges arise in F-theory constructions. After this, we turn to known constructions of LSTs with eight supercharges (i.e. minimal supersymmetry). This will also serve to illustrate how F-theory provides a single coherent framework for realizing LSTs.

\subsection{Theories with Sixteen Supercharges} \label{sec:sixteen}

To set the stage, we begin with little string theories with sixteen supercharges. In this case, we have two
possibilities given by $\mathcal{N} = (2,0)$ supersymmetry or $\mathcal{N} = (1,1)$ supersymmetry.
Note that only the former is possible in the context of 6D SCFTs.

One way to generate examples of $\mathcal{N}=(2,0)$ LSTs is to take $k$ M5-branes filling $\mathbb{R}^{5,1}$ and probing the
geometry $S_{\bot}^1 \times \mathbb{C}^2$ with $S_{\bot}^1$ a transverse circle of radius $R$. To reach the gravity-decoupling limit for an LST
we simultaneously send the radius $R \rightarrow 0$ and $M_{pl} \rightarrow \infty$  whilst holding fixed the effective string scale. In this case, it is the effective tension of an M2-brane wrapped over the circle which we need to keep fixed. Performing a reduction along this circle, we indeed reach type IIA string theory with $k$ NS5-branes. By a similar token, we can also consider IIB string theory with $k$ NS5-branes.
This realizes LSTs with $\mathcal{N}=(1,1)$ supersymmetry.

T-dualizing the $k$ NS5-branes of type IIA, we obtain type IIB\ string theory on the local geometry given by a configuration of $-2$ curves
arranged in the affine $\widehat{A}_{k-1}$ Dynkin diagram. Similarly, we also get an LST by taking
type IIA string theory on the same geometry.

Consider next the F-theory realization of these little string theories. First
of all, we reach the aforementioned theories by working with F-theory models
whose associated Calabi--Yau threefold takes the form
$T^{2}\times S$, in which $S\rightarrow C$ is an elliptically
fibered (non-compact) Calabi-Yau surface. If we treat the $T^{2}$ factor as the elliptic fiber of
F-theory, we get IIB\ on $S$, and if we treat the elliptic fiber of $S$ as the elliptic fiber of F-theory, we get (after shrinking the $T^{2}$
factor to small size) F-theory with base $T^2\times C$, which is dual to
IIA on $S$. To refer to both cases, it will be
helpful to label the auxiliary elliptic curve $T^2$ as $T_{F}^{2}$ (for fiber)
and the other elliptic curve as $T_{S}^{2}$ (since it lies in the surface $S$).

\begin{figure}[t!]
    \centering
        \includegraphics[scale=.8]{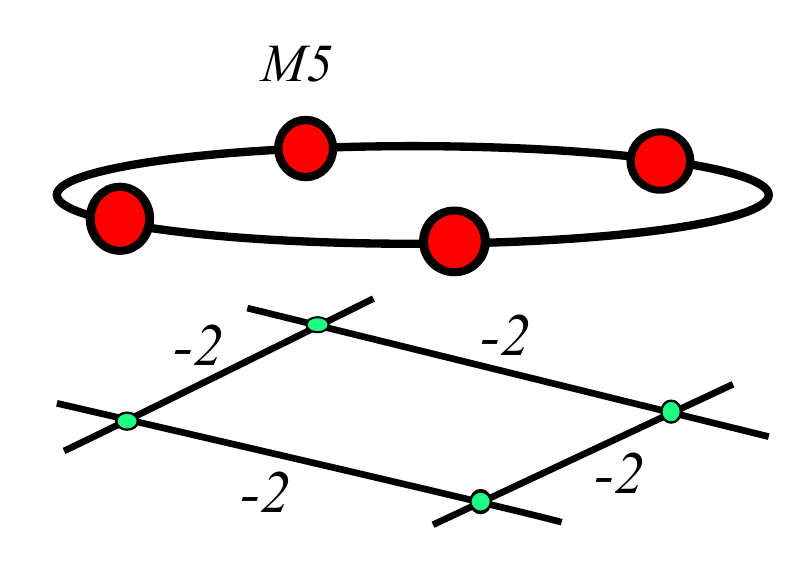}
            \caption{Depiction of the tensor branch of the $\mathcal{N}=(2,0)$ $\widehat{A}_{3}$ LST. In the top figure, we engineer this example using spacetime filling M5-branes probing the geometry $S_{\bot}^1 \times \mathbb{C}^2$. In the dual F-theory realization, we have four $-2$ curves in the base, which are arranged as the affine $\widehat{A}_3$ Dynkin diagram. The K\"ahler class of each $-2$ curve in the F-theory realization corresponds in the M-theory realization to the relative separation between the M5-branes.}
    \label{fig:M5LST}
\end{figure}

Now, by allowing $S$ to develop a singular elliptic fiber, we can
realize the same local geometries obtained perturbatively. For example, the
$\mathbb{C}^{2}/\mathbb{Z}_{k}$ lifts to a Kodaira
fiber of type $I_{k}$. Resolving this local singularity, we find $k$ compact cycles
$\Sigma_I \simeq \mathbb{P}^{1}$'s which intersect according to the affine $\widehat{A}_{k-1}$
Dynkin diagram. In this case the null divisor class is:
\begin{equation}
[\Sigma_{null}] = [\Sigma_1] + ...+ [\Sigma_{k}]
\end{equation}
that is, it is the ordinary minimal imaginary root of $\widehat{A}_{k-1}$.
By shrinking $T^2_F$ to small size, this engineers in F-theory
the $\mathcal{N}=(2,0)$ LST of $k$ M5-branes, or of $k$ NS5-branes in type IIA.
(See figure \ref{fig:M5LST} for a depiction of the A-type $\mathcal{N}=(2,0)$ LSTs.)
In the other case, one obtains F-theory on $T^2\times C$ whose fibers
have an $I_k$ singularity along $T^2\times \{0\}$.  Then $[\Sigma_{null}]$
is precisely the class of the F-theory fiber $T_S^2$, and supersymmetry enhances to $\mathcal{N} = (1,1)$.

More generally, we can consider any of the degenerations of
the elliptic fibration classified by Kodaira, i.e. the type $I_{n}%
,II,III,IV,I_{n}^{\ast},II^{\ast},III^{\ast},IV^{\ast}$ fibers and
produce a $\mathcal{N} = (2,0)$ model with that degeneration occuring as a curve configuration
on the F-theory base, as well as a $\mathcal{N} = (1,1)$ model with that same degeneration
occurring as the F-theory fiber over some $T^2$ in the base.

As a brief aside, a convenient way to realize examples of both the $\mathcal{N}=(1,1)$ and $\mathcal{N}=(2,0)$ theories
is to consider F-theory on the Schoen Calabi-Yau threefold $dP_{9}\times_{\mathbb{P}^{1}}dP_{9}$ \cite{schoen}.
Then, we can keep the elliptic fiber on one $dP_{9}$ factor generic, and allow the
other to degenerate. Switching the roles of the two fibers then moves us from
the IIA to IIB case. Note that although this strictly speaking only yields
eight real supercharges (as we are on a Calabi-Yau threefold), in the rigid
limit used to reach the little string theory, we expect a further enhancement
to either $\mathcal{N}=(2,0)$ or $\mathcal{N}=(1,1)$ supersymmetry. The specific chirality of the
supersymmetries depends on which elliptic curve we take to be in the base, and
which to be in the fiber of the corresponding F-theory compactification.

Let us also address whether each of the different Kodaira fiber
types leads us to a different little string theory. Indeed, some pairs
of Kodaira fiber types lead to identical gauge symmetries
in the effective field theory. To illustrate, consider the type $IV$ Kodaira
fiber, and compare it with the type $I_{3}$ fiber. There is a complex
structure deformation which moves the triple intersection appearing in the
type $IV$ case out to the more generic type $I_{3}$ case. This modulus,
however, is decoupled from the 6D little string theory. The reason is that if
we consider a further compactification on a circle, we reach a 5D gauge theory
which is the same for both fiber types. The additional complex structure
modulus from deforming $IV$ to $I_{3}$ does not couple to any of the modes of
the 5D theory. So, there does not appear to be any difference between
these theories. In other words, we should classify all of the $\mathcal{N}=(2,0)$
little string theories in terms of affine ADE Dynkin diagrams
rather than in terms of Kodaira fiber types.

For the $\mathcal{N} = (1,1)$ little string theories, the absence of a chiral structure actually leads to more possibilities. For example,
if we consider M-theory on an ADE singularity compactified on a further circle, we have the option of twisting by an outer automorphism of
the simply laced ADE Lie algebra \cite{Witten:1997kz}.
In other words, for the $\mathcal{N}=(1,1)$ theories we have an ABCDEFG classification according to all of the simple Lie algebras.

To realize these LSTs in F-theory, we make an orbifold of the previous
construction.  Suppose that $S\to C$ is
 an elliptically fibered (non-compact) Calabi--Yau surface which has
compatible actions of $\mathbb{Z}_m$ on the base and on the total space,
such that the action on the total space preserves the holomorphic $2$-form.
Then $\mathbb{Z}_m$ acts on $T_F^2\times S$ with the
action on $T^2$ being translation by a point of order $m$.  The quotient
$X:=(T_F^2 \times S)/\mathbb{Z}_m$ is then an elliptically fibered
Calabi--Yau threefold (with two genus one fibrations as before).

The elliptic fibration
$X \to (T_F^2 \times C)/\mathbb{Z}_m$ leads to an F-theory model
with $\mathcal{N}=(1,1)$ supersymmetry.
Note that the base $(T_F^2 \times C)/\mathbb{Z}_m$ of the F-theory fibration
contains a curve $\Sigma$ of genus $1$ and self-intersection $0$ such that $m\Sigma$
can be deformed into a one-parameter family although no smaller multiple can be
deformed.
Note also that if the action of $\mathbb{Z}_m$ on $S$ preserves the section of the fibration
$S\to C$, then $X\to (T_F^2 \times C)/\mathbb{Z}_m$ also has a section and,
as we will explain in section~\ref{sec:RANKZERO}, $m\in \{2,3,4,6\}$
since every elliptic fibration with section has a Weierstrass model
\cite{MR977771}.

There is a second fibration $X\to (S/\mathbb{Z}_m)$ which is a genus one
fibration without a section and leads to theories with $\mathcal{N}=(2,0)$ supersymmetry.
 We discuss additional details about
this second fibration, as well as  T-duality for these theories,
in section \ref{sec:Tduality}.

It is instructive to study the structure of the moduli space of the
LSTs with maximal supersymmetry. Recall that the tensor branch for
a $\mathcal{N}=(2,0)$ SCFT of ADE type $\mathfrak{g}$ is given by:
\begin{equation}\label{SCFTtwozero}
\mathcal{M}_{(2,0)}[\mathfrak{g}] = \mathbb{R}^{T}_{\geq 0} / \mathcal{W}_{\mathfrak{g}},
\end{equation}
where in the above, $T$ is the number of tensor multiplets and
$\mathcal{W}_{\mathfrak{g}}$ is the Weyl group of the ADE Lie algebra $\mathfrak{g}$. That is, the moduli space is given by a Weyl chamber of the ADE Lie algebra and is therefore non-compact. In the present case of LSTs, we see that the condition that we have a string scale leads to one further constraint on this moduli space, effectively ``compactifying'' it to the compact
Coxeter box for an affine root lattice \cite{Intriligator:1997dh}.

Finally, one of the prominent features of these examples is the manifest appearance of two elliptic fibrations in the
geometry. Indeed, in passing from the $\mathcal{N}=(2,0)$ theories to the $\mathcal{N}=(1,1)$ theories, we observe that we have simply switched the role of the two
fibrations. In section \ref{sec:Tduality}, we return to this general phenomenon for how T-duality of LSTs is realized in F-theory.

\subsection{Theories with Eight Supercharges}

Several examples of LSTs with minimal, i.e.
$(1,0)$ supersymmetry are realized by mild generalizations of the examples reviewed above.

To begin, let us consider again the case of $k$ coincident M5-branes filling $\mathbb{R}^{5,1}$ and probing the geometry
$S^1_{\bot} \times \mathbb{C}^2$. We arrive at a $(1,0)$ LST by instead taking a quotient of the $\mathbb{C}^2$ factor
by a non-trivial discrete subgroup $\Gamma_{G} \subset SU(2)$ so that the geometry probed by the M5-brane is $\mathbb{C}^2 / \Gamma_{G}$. The
discrete subgroups admit an ADE classification, and the corresponding simple Lie group $G_{ADE}$ specifies the gauge group factors on
the tensor branch. We reach a 6D\ SCFT by decompactifying the $S^{1}_{\bot}$. In this limit, we have an emergent $G_{L}\times G_{R}$ flavor
symmetry. From this perspective, the little string theory arises from gauging
a diagonal $G$ subgroup of the flavor symmetry. In the IIB realization of NS5-branes probing the affine geometry, applying S-duality takes
us to a stack of D5-branes probing an ADE singularity. On its tensor branch, this leads to an affine quiver gauge theory \cite{Douglas:1996sw}.

The F-theory realization of these LSTs is simply an affine A-type Dynkin diagram of $k$
curves of self-intersection $-2$ decorated with $I_{n},I^*_{n},IV^*,III^*,II^*$ fibers, respectively for $G=A_{n-1},D_{n+4},E_{6,7,8}$. We reach a 6D\ SCFT by decompactifying any of the $-2$ curves in the loop, and we recover a 6D SCFT with an emergent
$G_{L}\times G_{R}$ flavor symmetry. From this perspective, the little string theory arises from gauging
a diagonal $G$ subgroup of the flavor symmetry. Note that for $G\neq A_n$,
all these systems involve conformal matter in the sense of reference \cite{DelZotto:2014hpa}.

Another class of LSTs is given by taking $k$ M5-branes in heterotic M-theory, i.e. M-theory on $S^1 / \mathbb{Z}_{2} \times \mathbb{C}^2$.
In this case, we have two $E_8$ flavor symmetry factors; one for each endpoint of the interval $S^1 / \mathbb{Z}_2$. In this case, the gravity-decoupling limit requires us to collapse the size of the interval to zero size (i.e. to reach perturbative heterotic strings), whilst still holding the effective string scale finite. (The ratios of the lengths of subintervals between the endpoints and the various M5-branes to the length of the total interval will remain finite in the gravity-decoupling limit and provide parameters for the tensor branch.)
In perturbative heterotic string theory, we have $k$ NS5-branes probing $\mathbb{C}^2$. A related example is provided by instead working with the $Spin(32) / \mathbb{Z}_2$ heterotic string in the presence of $k$ NS5-branes. Indeed, once suitable Wilson line data has been specified, these two examples are T-dual to one another.

\begin{figure}[t!]
    \centering
        \includegraphics[scale=.8]{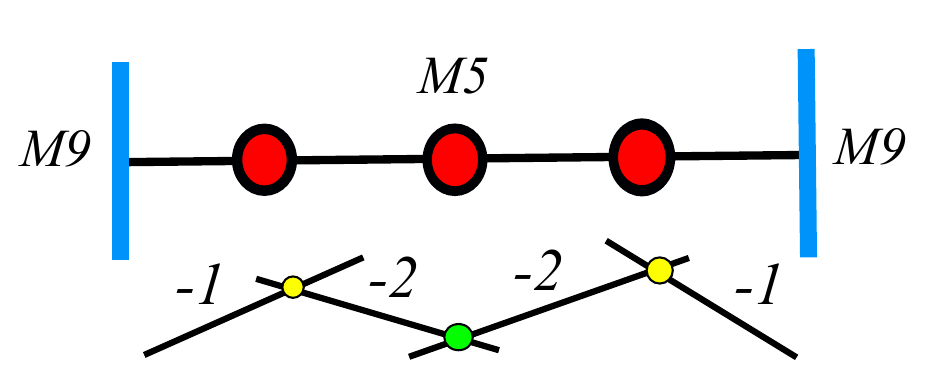}
            \caption{\textsc{Top}: Depiction of the LST realized by $k$ M5-branes in between the two Horava-Witten nine-brane walls of heterotic M-theory ($k=3$ above). This leads to an LST with an $E_8 \times E_8$ flavor symmetry \textsc{Bottom}: The Corresponding F-theory base given by the configuration of curves $[E_8],1,2,...,2,1,[E_8]$ for $k$ total compact curves. In this realization, the $E_8$ flavor symmetry is localized on two non-compact 7-branes, one intersecting each $-1$ curve.}
    \label{fig:M9LST}
\end{figure}

The F-theory realization of the theory of $k$ M5-branes is given by a non-compact base with a configuration of curves:
\begin{equation}
[E_{8}] \underset{k}{\underbrace{{1}%
,{2},...,{2},{1}}} [E_{8}]
\end{equation}
where we have indicated the flavor symmetry factors in square brackets. In this configuration, we reach the LST limit by holding fixed the volume of the null divisor (given by a sum over each divisor with multiplicity one), and collapse all other K\"ahler moduli to zero size. The construction of the T-dual characterization is somewhat more involved, and so we defer a full discussion to section \ref{sec:Tduality} and Appendix \ref{appx:Texamp}. See figure~\ref{fig:M9LST} for a depiction of the M-theory and F-theory realizations of this LST.

We can also combine the effects of different orbifold group actions. For example, we can consider $k$ M5-branes filling $\mathbb{R}^{5,1}$ and probing the geometry $S^1 / \mathbb{Z}_2 \times \mathbb{C}^{2} / \Gamma_{G}$. In F-theory terms, this is given by the geometry:
\begin{equation}
[E_{8}] \underset{k}{\underbrace{\overset{\mathfrak{g}}{1}%
,\overset{\mathfrak{g}}{2},...,\overset{\mathfrak{g}}{2},\overset{\mathfrak{g}%
}{1}}} [E_{8}]
\end{equation}
i.e. we decorate by a $\mathfrak{g}$-type ADE\ gauge symmetry over each curve
of self-intersection $-1$ or $-2$.  This geometry was studied in
detail in reference \cite{Aspinwall:1997ye}. Further blowups in the base are needed
for all fibers to remain in Kodaira-Tate form. This leads to conformal matter between each simply laced
gauge group factor \cite{DelZotto:2014hpa}.

Summarizing, we have seen in the above that the various LSTs which have been constructed via
perturbative string theory and M-theory all have a natural embedding in the
context of specific F-theory constructions. With this in mind, we now turn to a systematic
construction of all 6D LSTs in F-theory.

\section{Constraints from Tensor-Decoupling \label{sec:DECOUP}}

As a first step towards the classification of LSTs,
we now show how to classify possible bases using the ``tensor-decoupling criterion,'' that is,
the requirement that decoupling any tensor multiplet from an LST must take us to an SCFT.
In geometric terms, deleting any curve of the base (with possible fiber enhancements along this curve)
must take us back to an SCFT base (with possibly disconnected components).
Since all SCFTs have the structure of a tree-like graph of intersecting curves \cite{Heckman:2015bfa},
our task reduces to scanning over the list of \textit{connected}
SCFTs, and asking whether adding an additional curve
(with possible fiber enhancements on this curve) will produce an LST.
This inductive approach to classification will allow us to effectively
constrain the overall structure of bases for LSTs.

In this section we show how the tensor-decoupling criterion constrains many candidate bases for LSTs. We first
use this criterion to limit the possible graph topologies of curves in the base. Next, we give a general inductive rule for
how to take an SCFT and verify whether it enhances to an LST. We shall refer to this as an inductive classification, since it implicitly
accounts for all possible structures for LSTs. In section \ref{sec:ATOMIC} we use these constraints
to present a more explicit construction of possible bases for LSTs.

\subsection{Graph Topologies for LSTs}

For any compact curve $\Sigma$ in the base which remains in the gravity-decoupling limit,
the self-intersection $\Sigma^2$ must be $-n$ for $0 \leq n \leq 12$.  Moreover, since
having an F-theory model requires that $-4K$, $-6K$, and $-12K$ be  effective divisors,
if $K\cdot \Sigma + \Sigma^2 >0$ (so that $K\cdot \Sigma > 0$),
then $-4K$ would have multiplicity at least $4$ along
$\Sigma$, $-6K$ would have multiplicity at least $6$ along $\Sigma$, and $-12K$
would have multiplicity at least $12$ along $\Sigma$.  Since this is not allowed
in the Kodaira classification, we conclude that $2g-2 = K\cdot \Sigma + \Sigma^2\le0$,
in other words, that $\Sigma$ is
either $\mathbb{P}^1$ or $T^2$.
We now use the tensor-decoupling criterion to argue that the possible topologies of LST bases are limited to tree-like structures and appropriate degenerations of an elliptic curve.

Let us first show that a curve $\Sigma$ of self-intersection zero (of topology $\mathbb{P}^1$ or $T^2$) can only appear in isolation, i.e. it cannot intersect any other curve. If it met another curve and we decoupled everything that this curve touches, we would be left with an SCFT base containing a curve of self-intersection zero, a contradiction. If $\Sigma$ has genus $0$, then the base takes the form $\mathbb{C}\times \Sigma$, while if
$\Sigma$ has genus $1$, then the base takes the form
$(\mathbb{C}\times \Sigma)/\mathbb{Z}_m$, with $\mathbb{Z}_m$ acting on $\Sigma$ by
a translation and on $\mathbb{C}$ by multiplication by a root of unity.
Note that if either $g=0$ or $m=1$, the base is just a product.
Hence, to get a six-dimensional theory we must wrap seven-branes over $\Sigma$, i.e. we \textit{must} include a non-trivial fiber enhancement over this curve, unless $g=1$
and $m>1$.  (We will see examples of this latter case in section \ref{sec:Tduality}.)

Consider next adjacency matrices in which the off-diagonal entries are different from zero or one. For example, this can occur when a $-4$ and $-1$ curve form a closed loop (i.e. intersect twice), or when the same curves intersect along a higher order tangency. Again, this possibility is severely limited because if this were to a occur in a configuration with three or more curves, we would contradict the tensor-decoupling criterion. By the same token, the value of all off-diagonal entries are bounded below by two:
\begin{equation}
-2 \leq A_{IJ} \leq 0 \,\,\,\text{for}\,\,\, I \neq J,
\end{equation}
and in the case where $-2$ appears, we are limited to just two curves. The only possibilities for a rank one LST base (i.e. with two curves) are therefore:
\begin{equation}
1,1 \,\,\,\text{or}\,\,\, //2,2// \,\,\,\text{or}\,\,\, //4,1// \,\,\,\text{or}\,\,\, 2 || 2 \,\,\,\text{or}\,\,\, 4 || 1.
\end{equation}
In Appendix \ref{Appendix:Singularity} we analyze the possible fiber enhancements which can occur when the two curves meet along a tangency (i.e. do not respect normal crossing), as is the case in the last two configurations.

For all other LST bases, we see that all curves must be constructed from $\mathbb{P}^1$'s of self-intersection $-x$ for $1 \leq x \leq 12$, which all intersect with normal crossings, i.e. all off-diagonal entries of the adjacency matrix are either zero or one.

To further constrain the structure, we next observe that the base of any 6D SCFT is always tree-like \cite{Heckman:2013pva}. This means that the graph associated to an LST adjacency matrix can admit at most one loop, and when it contains a loop, there can be no additional curves branching off. This is because the tensor-decoupling criterion would be violated by joining a loop of curves to anything else. We are therefore left with two general types of configurations:

\smallskip

$\bullet$ Tree-like LSTs

$\bullet$ Loop-like LSTs.

Note that some of the tree-like structures we shall encounter can also be viewed as loops, that is, as degenerations of an elliptic curve.

\subsection{Inductive Classification}

To proceed further, we now present an inductive strategy for constructing the base of any LST with three or more curves. The main idea is that we simply need to sweep over the list of SCFT bases and ask whether we can append an additional curve of self-intersection $-y$ to such a base. By the remarks on decoupling already noted, we see that this additional curve can intersect either one curve or two curves of an SCFT base. In the latter case, we obtain a loop-like configuration of curves in the base. The latter possibility can only occur for an SCFT base which consists of a single line of curves (i.e. no branches emanating off of the primary spine of the base). The main condition we need to check is that after adding this curve, we obtain a positive semidefinite adjacency matrix. In particular, the determinant must vanish. Implicit in this construction is that we only append an additional curve compatible with the gluing rules for bases.

Consider first the case of an LST with adjacency matrix $A_{LST}$ which describes
a tree-like base given by adding a single curve of self-intersection $-y$ to some SCFT with adjacency matrix $A_{SCFT}$:
\be A_{LST}^{tree} = \left( \begin{array}{cccccccc}
y & -1 & 0 & 0 & ... & 0& 0& 0  \\
-1 &   &  &  &   &  &  &   \\
0 &  &  &  & A_{SCFT} & & &   \\
\vdots &&&& &&& \vdots \\
0 &  &  &  & &  & &   \\
 \end{array} \right)
\ee
Let $A_{SCFT'}$ be the matrix obtained from $A_{SCFT}$ by removing the 1st column and the 1st row. Evaluating the determinant of $A_{LST}^{tree}$, we
obtain the condition:
\be
0 = \text{det} (A_{LST}^{tree}) = y \, \text{det}(A_{SCFT}) - \text{det}(A_{SCFT^\prime}).
\ee
or:
\be
y = \text{det}(A_{SCFT^\prime})/\text{det}(A_{SCFT}).
\ee

Consider now the case of a loop-like LST. In this case, the only SCFTs we need consider are those constructed from a single line of curves (i.e. no trivalent vertices at all), and we can only add the additional curve to the leftmost and rightmost ends of a candidate SCFT. The adjacency matrix is then of the form:
\be A^{loop}_{LST} = \left( \begin{array}{cccccccc}
y & -1 & 0 & 0 & ... & 0& 0& -1  \\
-1 &   &  &  &   &  &  &   \\
0 &  &  &  & & & &   \\
\vdots &&&& A_{SCFT} &&& \vdots \\
0 &  &  &  &  & & &\\
-1 &  &  &  & &  & &   \\
 \end{array} \right)
\ee
To have an LST we must have
\be
0 = y \, \text{det}(A_{SCFT}) - (A_{SCFT})_{(1,1)} - (A_{SCFT})_{(N-1,N-1)} + 2 (-1)^{N+1} (A_{SCFT})_{(1,N-1)}
\ee
where we have denoted the $(i,j)^{th}$ minor of $A_{SCFT}$ by an appropriate subscript. Solving for $y$, we obtain:
\be
y = \frac{ (A_{SCFT})_{(1,1)} + (A_{SCFT})_{(N-1,N-1)} - 2 (-1)^{N+1} (A_{SCFT})_{(1,N-1)} }{\text{det}(A_{SCFT})}.
\ee

The above algorithm allows us to systematically classify LSTs: From this structure, we see that the locations of where we can add an additional curve to an existing SCFT are quite constrained. Indeed, in order to \textit{not} produce another SCFT, but instead an LST, we will typically only be able to add our extra curve at the end of a configuration of curves, or at the second to last curve. Otherwise, we could not reach an SCFT upon decoupling other curves in the base.

\subsection{Low Rank Examples}

\begin{table}
\begin{center}
\begin{tabular}{|l|ccccccccccccc|}
\hline
\phantom{$\Bigg|$}SCFT&12& 13&14&15&16&17&18&19&1(10)&1(11)&1(12)&22&23\\
\hline
\phantom{$\Big|$}$\text{det}(A)$&1&2&3&4&5&6&7&8&9&10&11&3&5\\
\hline
\phantom{$\Big|$}$y$ & 5 & 3& 7/3 & 2 & 9/5 & 5/3 & 11/7 & 3/2 & 13/9 & 7/5 & 15/11 & 2 & 7/5\\
\hline
\end{tabular}
\vspace{3mm}
\caption{Candidate loop-like rank two LSTs from adding an additional curve to a rank two SCFT.}\label{rank2decouplingalgo}
\end{center}
\end{table}

To illustrate how the algorithm works in practice, we now give some low rank examples. In table \ref{rank2decouplingalgo} we list all of the rank two SCFT bases which we attempt to enhance to loop-like LST. Of the cases where $y$ is an integer, some are further eliminated since the resulting base requires further blowups.\footnote{For example, configurations
such as $//512//$ and $//313//$ require a further blowup.
Doing this, we instead reach a four curve LST base, respectively
given by $//6131//$ and $//4141//$.} The full list of rank two LST bases is then:
\be
\begin{gathered}
\text{Three curve LST bases:}\\
2\overset{2}{\bigtriangledown}2\,\,\,\text{and}\,\,\,121\,\,\,\text{and}\,\,\,212\,\,\,\text{and}\,\,\,//222//
\end{gathered}
\ee
where the first entry denotes a triple intersection of $-2$ curves (that is, a type $IV$ Kodaira degeneration),
and $//x_1x_2...x_{n+1}//$ denotes a loop in which the two sides are identified.

\section{Atomic Classification of Bases \label{sec:ATOMIC}}

In principle, the remarks of the previous section provide an implicit way to characterize all LSTs. Indeed, we simply need to sweep over the list of bases for SCFTs obtained in reference \cite{Heckman:2015bfa} and then determine whether there is any place to add one additional curve to reach an LST. The self-intersection of this new curve is constrained by the condition that the determinant of the adjacency matrix vanishes, and the location of where we add this curve is likewise constrained by the tensor-decoupling criterion.

In this section we use the atomic classification of 6D\ SCFTs presented in
\cite{Heckman:2015bfa} to perform a corresponding atomic classification of bases for LSTs.
We now use the explicit structure of 6D\ SCFTs found in \cite{Heckman:2015bfa} to further
cut down the possibilities. It is helpful to
view the bases as built out of smaller \textquotedblleft
atoms\textquotedblright\ and \textquotedblleft radicals\textquotedblright. In
particular, we introduce the convention of a \textquotedblleft
node\textquotedblright\ referring to a single curve in which the minimal fiber
type leads to a D or E-type gauge algebra. We refer to a \textquotedblleft
link\textquotedblright\ as any collection of curves which does not contain any
D or E-type gauge algebras for the minimal fiber type. The results of \cite{Heckman:2015bfa}
amount to a classification of all possible links, as well as all possible ways
of attaching links to the nodes. Quite remarkably, the general structure of
the resulting bases is quite constrained. For all 6D\ SCFTs, we can filter the
theories according to the number of nodes in the graph. These nodes are always
arranged along a single line joined by links:%
\begin{equation}
S_{0,1}\overset{S_{1}}{g_{1}}L_{12}\overset{\mathbf{I}^{\oplus s}}{g_{2}%
}L_{2,3}g_{3}...g_{k-2}L_{k-2,k-1}\overset{\mathbf{I}^{\oplus t}}{g_{k-1}%
}L_{k-1,k}\overset{\mathbf{I}^{\oplus u}}{g_{k}}S_{k,k+1},\label{basequiver}%
\end{equation}
here, the $g_{i}$'s denote the nodes, the $L_{i,i+1}$'s denote interior links
(since they join to two nodes) and the $S$'s are side links as they can only
join to one node. The notation $\mathbf{I}^{\oplus m}$ refers to decorating by $m$
small instantons, these are further classified according to partitions of $m$
(i.e. how many of the small instantons are coincident with one another). One
of the key points is that for $k\geq6$, there is no decoration on any of the
interior nodes, i.e. for $3\leq i\leq k-2$. This holds both for the types of
links which can attach to these nodes (which are always the minimal ones
forced by the resolution algorithm of reference \cite{Heckman:2013pva}), as well as the
possible fiber enhancements (there are none). When $k=5$, it is possible to
decorate the middle node $g_{3}$ by a single $-1$ curve. In reference
\cite{Heckman:2015bfa}, the explicit form of all such sequences of $g$'s, as well as the
possible side links and minimal links was classified. An additional important
property is that all of the interior links blow down to a trivial endpoint,
the blowdown of a single $-1$ curve.

Turning now to LSTs, we can ask whether we can add one more curve to the base
quiver, resulting in yet another tree-like graph, or in a loop-like graph. By
inspection, we can either add this additional curve to a side link, an
interior link, or a base node.

\subparagraph{Restrictions on Loop-like Graphs:}

In fact, a general loop-like graph which is an LST is tightly constrained by
the tensor-decoupling criterion. The reason is that if we consider the resulting
sequence of nodes, we must have a pattern of the form:%
\begin{equation}\label{LoopConfMat}
//g_{1}L_{12}g_{2}L_{2,3}g_{3}...g_{k-2}L_{k-2,k-1}g_{k-1}L_{k-1,k}%
g_{k}L_{k,1}//
\end{equation}
where the notation \textquotedblleft$//$\textquotedblright\ indicates that the
left and the right of the base quiver are joined together to form a loop. Now,
another important constraint from reference \cite{Heckman:2015bfa} is that the minimal fiber
type on the nodes obeys a nested sequence of containment relations. But in a
loop, no such ordering is possible. We therefore conclude that all of the
nodes for a loop-like LST must be identical, and moreover, that the interior
links must all be minimal. We therefore can specify all such loops simply by
the type of node (i.e. a $-4$ curve, a $-6$ curve, a $-8$ curve or a $-12$
curve), and the number of such nodes.

For this reason, we now confine our attention to tree-like graphs, i.e.
where we add an additional curve which intersects only one other curve in the base.
The main restriction we now derive is that the resulting configuration
of curves is basically the same as that of line (\ref{basequiver}). Indeed, we
will simply need to impose further restrictions on the possible side links and
sequences of nodes which can appear in an LST base.

\subparagraph{Restrictions on Adding to Interior Links:}

Our first claim is that we can only possibly add an extra curve to an interior
link in a base with two or fewer nodes. Indeed, suppose to the contrary.
Then, we will encounter a configuration such as:%
\begin{equation}
...g_{i}\overset{y}{L_{i,i+1}}g_{i+1}...
\end{equation}
where $y$ denotes our additional curve attached in some way to the link. The
notation \textquotedblleft...\textquotedblright\ denotes the fact that there
is at least one more curve in the base. Now, since the interior link blows
down to a single $-1$ curve, we will get a violation of normal crossing. This
is problematic if we have one additional curve (as denoted by the
\textquotedblleft...\textquotedblright), since deleting that curve would
produce a putative SCFT with a violation of normal crossing, a contradiction.
By the same token, in a two node base, if any side links are attached to this
node, then we cannot add anything to the interior link. This leaves us with
the case of just:%
\begin{equation}
g_{1}\overset{y}{L_{1,2}}g_{2}.
\end{equation}
In this case, it is helpful to simply enumerate once again all of the possible
interior links, and ask whether we can attach an additional curve. This we do
in Appendix \ref{appx:nonDEbases}, finding that the options are severely limited. Summarizing,
then, we find that we can attach an extra curve to an interior link only in
the case where there are two nodes, and then only if these two nodes do not
attach to any side links.

\subparagraph{Restrictions on Adding to Nodes:}

Let us next turn to restrictions on adding an extra curve to a node in a base.
If we add a curve to a node, we observe that this extra curve must have
self-intersection $-1$. Note that the endpoint of the SCFT must therefore be
trivial in these cases. We now ask which of the nodes of the base can support
an additional $-1$ curve. Since we must be able to delete a single curve and
reach a collection of SCFTs, we cannot place this $-1$ curve too far into the
interior of the configuration. More precisely, we see that for $k\geq7$ nodes,
we are limited to adding a $-1$ curve to the first three, or last three
nodes.\ In the specific case where we attach a $-1$ curve to the third
interior node, we see that there cannot be any side links whatsoever.
Otherwise, we would find a subconfiguration of curves which
is not a 6D\ SCFT.

\subparagraph{Restrictions on Adding to Side Links:}

Consider next restrictions on adding an extra curve to a side link. In the
case of a small instanton link such as $1,2...,2$, we can append an additional
$-1$ curve to the rightmost $-2$ curve, but then it can no longer function as
a side link (via the tensor-decoupling criterion). In Appendices D and E we determine the
full list of LSTs comprised of just adding one more curve to a side link. If
we instead attempt to take an existing SCFT and add an additional curve to a
side link to reach an LST, then we either produce a new side link (i.e. if
the curve has self-intersection $-1,-2,-3$ or $-5$), or we produce a base
quiver with one additional node (i.e. if the curve has self-intersection
$-4,-6,-7,-8,-9,-10,-11,-12$). Phrased in this way, we see that the rules for
which side links can join to an SCFT are slightly different, but cannot alter
the overall topology of a base quiver from the case of an SCFT.

Summarizing, we see that unless we have precisely two nodes, and no side
links, we cannot decorate any interior link. Moreover, we can only decorate
the three leftmost and rightmost node in special circumstances. So in other
words, the general structure of a tree-like LST base is essentially the same
as that of a certain class of SCFTs. All that remains is for us to determine
the possible sequences of nodes (with no decorations) which can generate an
LST, and to also determine which of our side links can be attached to an SCFT
such that the resulting configuration is an LST.

\subparagraph{Overview of Appendices:}

This final point is addressed in a set of Appendices. In the
appendices we collect a full list of the building blocks for constructing
LSTs. The tensor-decoupling criterion prevents a direct gluing of smaller LSTs
to reach another LST. Rather, we are always supplementing
an SCFT to reach an LST. Along these lines, in Appendix \ref{appx:DEbases} we collect the list of bases which are comprised of a single spine of nodes with no further decoration from side links. In Appendix \ref{appx:nonDEbases}
we collect the full list of bases in which no nodes appear. Borrowing from the terminology used for 6D SCFTs, these links are ``noble'' in the sense that they cannot attach to anything else in the base. Finally, in Appendix \ref{appx:NOVEL} we give a list of LSTs given by attaching a single side link to a single node. Much as in the classification of 6D SCFTs,
the further task of sweeping over all possible ways to decorate a base
quiver by side links is left implicit (as dictated by the number of blowdowns induced by a given side link). All of these rules follow directly
from reference \cite{Heckman:2015bfa}.

This completes the classification of bases for LSTs. We now turn to the classification of elliptic fibrations over a given base.

\section{Classifying Fibers} \label{sec:FIBERS}

Holding fixed the choice of base, we now ask whether we can enhance the singularities over curves of the base whilst keeping
all fibers in Kodaira-Tate form. As this is a purely local question (i.e. compatible with the matter enhancements over the
neighboring curves), most of the rules for adding extra gauge groups / matter are fully
specified by the rules spelled out in reference \cite{Heckman:2015bfa}. Rather than repeat this discussion, we refer the interested
reader to these cases for further discussion of the ``standard'' fiber enhancement rules for curves which intersect with normal crossings.

There are, however, a few cases which cannot be understood using just the SCFT considerations of reference \cite{Heckman:2015bfa}. Indeed, we have already seen that a curve of self-intersection zero, an elliptic curve, tangent intersections and triple intersections of curves can all
occur in the base of an LST. We have also seen, however, that all of these cases are comparatively
``rare'' in the sense that they do not attach to larger structures. Our plan in this section will therefore be to deal with all of these low
rank examples. In Appendix \ref{anomalyagainandagain} we give general constraints from anomaly cancellation in F-theory models and in
Appendix \ref{Appendix:Singularity} we present some additional technical material on matter content in the case of singular curves in the base. Finally, compared with the case of 6D SCFTs, the available fiber enhancements over a given base
are also comparatively rare. To illustrate this point, we give some examples in which the base is an affine Dynkin diagram of $-2$ curves. In these cases, the presence of the additional imaginary root (and the constraints from anomaly cancellation) typically dictate a small class of possible fiber enhancements.

\subsection{Low Rank LSTs}

In this subsection we give a complete characterization of fiber enhancements for low rank LSTs. To begin, we consider the case of the rank zero LSTs, i.e. those where the F-theory base consists of a single compact curve. In these cases, we \textit{only}
get a 6D theory once we wrap some 7-branes over the curve
unless the normal bundle of the curve is a torsion line bundle.
An interesting feature of this and related examples is that because the corresponding tensor multiplet is non-dynamical it cannot participate in the Green-Schwarz mechanism and we must cancel the anomaly using just the content of the gauge theory sector. We then turn to the other low rank examples where other violations of normal crossing appear. In all of these cases, the F-theory geometry provides a systematic tool for determining which of these structures can embed in a UV complete LST.

\subsubsection{Rank Zero LSTs} \label{sec:RANKZERO}

In a rank zero LST, we have a single compact curve, which must necessarily have self-intersection zero. There are only a few inequivalent configurations consisting of a single curve with self-intersection zero:
\be\label{LSTrankzero}
\Sigma_{null} = P, I_0, I_1, II, {}_mI_0, {}_mI_1.
\ee
Here $I_0$ (resp. $P$) is shorthand for a base B consisting of a smooth torus (resp. two sphere) with trivial normal bundle $T^2 \times \mathbb{C}$ (respectively $\mathbb{P}^1\times \mathbb{C}$), while $I_1$ (resp. $II$) is a curve with a node (resp. a cusp) singularity and trivial normal bundle. These configurations can give rise to LSTs \emph{only if} 7-branes wrap $\Sigma_{null}$. Otherwise, we do not have a genuine 6D model. The variants ${}_mI_0$
and ${}_mI_1$ describe curves whose normal bundle is torsion of order $m>1$;
these can also support 6D theories for $m\in\{2,3,4,6\}$ as
discussed below.  Observe also
that if we apply the tensor-decoupling criterion in these cases, we find that the resulting 6D SCFT is empty, i.e. trivial.

As curves of self-intersection zero do not show up in 6D SCFTs, it is important to explicitly list the possible singular fiber types which can arise on each curve of line \eqref{LSTrankzero}.\footnote{In what follows we focus on those cases where the fiber enhancement leads to a non-abelian gauge symmetry, i.e. a gauge theory description. In the cases where we have a type $I_1$ or type $II$ fiber enhancement, the resulting 6D theory will consist of  some number of weakly coupled free hypermultiplets, where the precise number depends on whether the base curve has non-trivial arithmetic and / or geometric genus. Much as in the case of 6D SCFTs, these cases can be covered through a mild extension of the analysis presented in reference \cite{Heckman:2015bfa}.
See also \cite{MPT} for additional information about these theories.}

We begin with the ``multiple fiber phenomenon'' -- a genus one curve $\Sigma$
whose normal bundle is torsion of order $m>1$.
The F-theory base $B$ is a (rescaled) small neighborhood of
$\Sigma$, and its canonical bundle $\mathcal{O}_B(K_B)$
must also be torsion of the same
order by the adjunction formula.  Now to construct a Weierstrass model,
we need sections $f$ and $g$ of $\mathcal{O}_B(-4K_B)$ and
$\mathcal{O}_B(-6K_B)$,
respectively, but nontrivial torsion bundles do not have nonzero sections.
Thus, in order to have a nonzero $f$, the order $m$ of the torsion must
divide $4$, while to have a nonzero $g$, the order $m$ must divide $6$.
There are thus three cases:
\begin{enumerate}
\item If $m=2$, then both $f$ and $g$ may be nonzero.
\item If $m=3$ or $6$, then $f$ must be zero but $g$ may be nonzero.
\item If $m=4$, then $g$ must be zero but $f$ may be nonzero.
\end{enumerate}
For any other value of $m>1$, Weierstrass models do not exist (since $f$
and $g$ are not both allowed to vanish identically).

Note that the fact that some quantities obtained from coefficients in
a Weierstrass model are sections of torsion bundles also provides the
possibility that those sections do not exist (if they are known to be
nonzero).  As described in
Table~4 of \cite{Grassi:2011hq}, the criterion for deciding whether
a given Kodaira type leads to a gauge algebra whose Dynkin diagram
is simply laced or not simply laced reduces in almost every case
to a question of whether a certain quantity has a square
root.\footnote{In the remaining case, one must consider a more complicated
cubic equation, but in the situation being described here, the question
in that case boils down to the existence or non-existence of a cube
root.  If the bundle of which the desired cube root is a section is
a $3$-torsion bundle, the cube root cannot exist.}
If the desired square root is in fact a section of a
$2$-torsion bundle, then it cannot exist.

We can give explicit examples of this phenomenon which do not involve
enhanced gauge symmetry, using the framework outlined in
section~\ref{sec:sixteen}.  We  start with $S=T_S^2\times \mathbb{C}$, where
$T_S^2$ admits an automorphism of order $m$ which acts faithfully on
the holomorphic $1$-form. If we extend the action to include multiplication
by an appropriate root of unity on $\mathbb{C}$, then the holomorphic $2$-form on $S$ is
preserved.  As is well-known, such automorphisms exist
exactly for $m\in\{2,3,4,6\}$.  As explained in
section~\ref{sec:sixteen}, the quotient $(T_F^2\times S)/\mathbb{Z}_m$
has an elliptic fibration
$(T_F^2\times S)/\mathbb{Z}_m \to (T_F^2\times\mathbb{C})/\mathbb{Z}_m$ whose
fibers over  $T_F^2\times\{0\}$  are all nonsingular elliptic curves,
but with $\mathbb{Z}_m$ acting upon them as loops are traversed on
$T^2_F$.  This same geometry has a second fibration
$(T_F^2\times S)/\mathbb{Z}_m \to S/\mathbb{Z}_m$ with no section and some
multiple fibers in codimension two, which will be further discussed in section \ref{sec:Tduality}.

Turning now to enhancements of fibers,
from the relation between anomaly cancellation and enhanced singular fibers \cite{Sadov:1996zm,Grassi:2000we,Grassi:2011hq,Morrison:2011mb} we find all possible gauge theories compatible with a given choice of base curve. This is summarized in table \ref{rankzeroLSTs}. We find that in general, such theories can support hypermultiplets in the adjoint representation (denoted $Adj$), the two-index symmetric representation (denoted $\text{sym}$) and the $n$-index anti-symmetric representation (denoted $\Lambda^{n}$).
\begin{table}
\begin{center}
\begin{tabular}{l c}
\phantom{\Bigg|}Base Curve & Matter Content\\
\hline
\phantom{\Big|}$I_0$& Any simple Lie algebra, $n_{Adj} =1$\\
\hline
\phantom{\Big|}$I_1$,$II$& $\mf{su}(N)$, $n_{\rm sym} = 1$, $n_{\Lambda^2}=1$\\
& $\mf{su}(6)$, $n_{f} = 1$, $n_{\rm sym} = 1$, $n_{\Lambda^3}= \frac{1}{2}$\\
\hline
\phantom{\Big|}$P$& $\mathfrak{su}(N)$, $N \geq2$, $n_{f} = 16$, $n_{\Lambda^{2}}=2$\\
\phantom{\Big|}& $\mathfrak{su}(6)$, $n_{f} = 17$, $n_{\Lambda^{2}}=1$, $n_{\Lambda^{3}}= \frac{1}{2}$\\
\phantom{\Big|}& $\mathfrak{su}(6)$, $n_{f} = 18$, $n_{\Lambda^{3}}= 1$\\
\phantom{\Big|}& $\mathfrak{sp}(N)$, $N \geq1$, $n_{f} = 16$, $n_{\Lambda^2} = 1$\\
\phantom{\Big|}& $\mathfrak{sp}(3)$, $n_{f} = 17\frac{1}{2}$, $n_{\Lambda^3} = \frac{1}{2}$\\
\phantom{\Big|}& $\mathfrak{so}(N)$, $N = 6,...,14$, $n_{f} = N-4$, $n_{s} = \frac{64}{d_{s}}$\\
\phantom{\Big|}& $\mathfrak{g}_{2}$, $n_{f} =10 $\\
\phantom{\Big|}& $\mathfrak{f}_{4} $, $n_{f} = 5$\\
\phantom{\Big|}& $\mathfrak{e}_{6}$, $n_{f} = 6$\\
\phantom{\Big|}& $\mathfrak{e}_{7}$, $n_{f}=4$\\
\phantom{\Big|}& $\mathfrak{e}_{8}$, $n_{\text{inst}}=12$\\
\hline
\end{tabular}
\vspace{5mm}
\caption{ Rank zero LSTs. In the above, $Adj$ refers to the adjoint representation,
$\text{sym}$ refers to a two-index symmetric representation, and $\Lambda^{n}$ refers to an $n$-index anti-symmetric representation.}\label{rankzeroLSTs}
\end{center}
\end{table}

The greatest novelty here relative to the case of 6D SCFTs are the theories with $n_{Adj}=1$ or $\mf{su}(N)$ gauge algebra, $n_{\rm sym} = 1$, $n_{\Lambda^2}=1$ or, in the special case of $\mf{su}(6)$, $n_f =1$, $n_{\rm sym} = 1$, $n_{\Lambda^3}=1/2$.  The first of these cases, with $n_{Adj}=1$, corresponds simply to a smooth curve of genus 1 in the base.  The cases with symmetric representations of $\mf{su}(N)$, on the other hand, arise when the \emph{base} curve is of Kodaira type $I_1$ (i.e. has a nodal singularity).  As reviewed in Appendix \ref{Appendix:Singularity}, the notion of genus is ambiguous for singular curves.  A type $I_1$ curve has topological genus 0 but arithmetic genus 1, and as a result it must support a hypermultiplet in the two-index symmetric representation, rather than one in the adjoint representation.

For LSTs, these are the only examples in which a curve of (arithmetic) genus 1 shows up, and a curve of genus $g \geq 2$ is never allowed.  Note also that a curve of genus $0$ and self-intersection $0$ cannot itself support an $\mf{e}_8$ gauge algebra: it must be blown up at 12 points, resulting in an $\mf{e}_8$ theory with 12 small instantons:
\begin{equation}
(12),1,2,2,2,2,2,2,2,2,2,2,2
\end{equation}

\subsubsection{Rank One LSTs}\label{rankoneLSTs}

Consider next the case of rank one LSTs, i.e. those in which there are two compact curves in the base. As we have already remarked,
in this and all higher rank LSTs, all the curves of the base will be $\mathbb{P}^1$'s, and moreover, they will have self-intersection
$-x$ for $1 \leq x \leq 12$. Now, in the case of two curves, we can have various violations of normal crossing. For example, we can have curves which intersect along a tangency which occurs in $4 || 1$.
Observe that in this cases, there is a smoothing deformation which takes us from an order two tangency to a loop, i.e. we can deform to $// 4 , 1 //$. In addition to these rank one LSTs, there are two more configurations given by $1,1$ and $// 2 , 2 //$.  The former configuration $1,1$ is in some sense the most ``conventional'' possibility (as all intersections respect normal crossing).

To this end, let us first discuss fiber enhancements for the $1,1$ configuration. We shall then turn to the cases where there is either a violation of normal crossing or a loop configuration. Whenever curves with gauge algebras intersect, matter charged under each gauge algebra will pair up into a mixed representation of the gauge algebras.  The mixed anomaly condition places strong constraints on which representations are allowed to pair up.  The allowed set of mixed representations for two curves intersecting at a single point is given in section 6.2 of reference \cite{Heckman:2015bfa}. Consider for example, the $1,1$ base. We have:
\begin{equation}
\overset{\mf{g}_L}1 \,\,    \overset{\mf{g}_R}1
\end{equation}
with the following list of allowed gauge algebras:
\begin{itemize}
\item $\mf{g}_L = \mf{so}(M)$, $\mf{g}_R = \mf{sp}(N)$, $M = 7,...,12$, $M -5 \geq N$, $4 N + 16 \geq M$.
\item $\mf{g}_L = \mf{so}(M)$, $\mf{g}_R = \mf{sp}(N)$, $M = 7$, $N \leq 6$.
\item $\mf{g}_L = \mf{g}_2$, $\mf{g}_R = \mf{sp}(N)$, $N \leq 7$.
\item $\mf{g}_L = \mf{sp}(M)$, $\mf{g}_R = \mf{sp}(N)$, $2M + 8 \geq 2N$, $2N + 8 \geq 2M$.
\item $\mf{g}_L = \mf{sp}(M)$, $\mf{g}_R = \mf{su}(N)$, $2M + 8 \geq N$, $N + 8 + \delta_{N,3}+ \delta_{N,6} \geq 2M$.
\item $\mf{g}_L = \mf{su}(M)$, $\mf{g}_R = \mf{su}(N)$, $M + 8 + \delta_{M,3}+ \delta_{M,6} \geq N$, $N + 8 + \delta_{N,3}+ \delta_{N,6} \geq M$.
\item $\mf{g}_L = \mf{f}_4$, $\mf{e}_6$, $\mf{e}_7$  or $\mf{e}_8$, $\mf{g}_R = \emptyset$.
\end{itemize}
Here, it is understood that $\mf{sp}(0)$ is the same as an empty $-1$ curve, and $\mf{e}_8$ on a $-1$ curve implies that 11 points on the $-1$ curve have been blown up.

Consider next the configuration $//2, 2//$, i.e. a loop of two $-2$ curves. In this case
the only gauge algebra enhancement is given by
\be
//\overset{\mf{su}(N)}2 \, \,\overset{\mf{su}(N)}2//
\ee
with a bifundamental localized at each intersection point.

Finally, we turn to the case of the bases $//4 , 1 //$ and $4 || 1$. In both cases, the only enhancement possible is an SO-type algebra over the $-4$ curve and an $Sp$-type or SU-type algebra over the $-1$ curve. The matter content of the theory, however, depends on the type of intersection. Consider first the case of
\be
// \overset{\mf{so}(2N+8)}4\,\,\overset{\mf{sp}(N)}1 //
\label{eq:onefour}
\ee
with a half hypermultiplet localized at each intersection point. Indeed, we can reach this theory by starting from the 6D SCFT:
\be
[ \mathfrak{so}(2N+8) ]\,\,\overset{\mf{sp}(N)}1 \,\, [ \mathfrak{so}(2N+8) ]
\ee
and gauging the diagonal subalgebra of the flavor symmetry.

In the case of the tangential intersection $4 || 1$, we again find novel configurations of matter which are missing from the case of normal crossing.  First,
we can consider a configuration in which the gauge algebras are the same as those of (\ref{eq:onefour}):
\begin{equation}
 \overset{\mf{so}(2N+8)}4\,||\,\overset{\mf{sp}(N)}1
\end{equation}
However, there is now a single full hypermultiplet in the bifundamental of the two gauge algebras rather than two half hypermultiplets.
This configuration cannot be realized in F-theory, since the tangential
intersection removes the monodromy from the $I_{2N}$ locus leading to gauge
algebra $\mf{su}(2N)$ instead.  It would be interesting to find a
field-theoretic reason for excluding this case.

Second, there is a similar configuration with a unitary rather than 
symplectic gauge algebra:
\begin{equation}
 \overset{\mf{so}(2N+8)}4\,||\,\overset{\mf{su}(2N)}1
\end{equation}
with a hypermultiplet in the bifundamental of the two gauge algebras, as well as a hypermultiplet in the two index anti-symmetric
representation of the $\mathfrak{su}(n)$ factor, all of which are located at the collision point between the two branes.
This configuration can be realized in F-theory.

\subsubsection{Rank Two LSTs}\label{ranktwoLSTs}

We now turn to the case of rank two LSTs, i.e. those with three curves.
Here, we can have no tangential intersections. In this case, the adjacency matrix
again provides only partial information about the geometry of intersecting curves. In the case where we have normal crossings
for all pairwise intersections, the rules of enhancing fiber enhancements follow from reference \cite{Heckman:2015bfa} and
are also reviewed in Appendix \ref{anomalyagainandagain}. There is, however, also the possibility of a Kodaira type $IV$ configuration of $-2$ curves, i.e. $2\overset{2}{\bigtriangledown}2$. We shall therefore confine our attention to fiber decorations with this base.

To illustrate, suppose the gauge algebra localized on each of the three $-2$ curves is $\mf{su}(2)_i, i=1,2,3$. Anomaly cancellation dictates that in such a case there must be a single half-trifundamental $\frac{1}{2}(\textbf{2}, \textbf{2}, \textbf{2})$ plus two fundamentals charged under each $\mf{su}(2)$. We note in passing that the loop-like configuration $//222//$ also admits a similar enhancement in the gauge algebras, i.e. with $\mf{su}(2)_i, i=1,2,3$ gauge groups, but that the corresponding matter content is given by three bifundamentals $(\textbf{2}, \textbf{2}, \textbf{1})$, $(\textbf{2}, \textbf{1}, \textbf{2})$, $(\textbf{1}, \textbf{2}, \textbf{2})$. These two configurations have the same anomaly polynomials.

But in contrast to the $//222//$ configuration, for $2\overset{2}{\bigtriangledown}2$,
no other gauge algebra enhancements are possible. To see this,
suppose we $\mathfrak{su}(N)$ factors on each $-2$ curve. We would then need $N^2$ fundamentals of each $\mathfrak{su}(N)$
to get a $(N,N,N)$ representation. However, anomaly cancellation considerations constrain us to 2N such fundamentals. In other words, we are limited to $N \leq 2$.

\subsection{Higher Rank LSTs}\label{rankhigherLSTs}

Turning next to the case of LSTs with at least four curves in the base, all of these local violations of normal crossing do not appear. Nevertheless, we encounter such violations when we attempt to blow down $-1$ curves which touch more than two curves. Even so, the local rules for fiber enhancements follow the same algorithm already spelled out in detail in reference \cite{Heckman:2015bfa}. In some cases, however, there can be additional restrictions compared with the fiber enhancements which are possible
for 6D SCFTs. To illustrate, we primarily focus on some simple examples, i.e. the affine ADE bases, fiber decorations for the
base $1,2,...,2,1$, and fiber decorations for the loop-like bases.

\subsubsection{Affine ADE Bases}

Let us next consider the case of fiber enhancements in which the base is given by an affine Dynkin diagram of $-2$ curves. If we assume that no further blowups are introduced in the base, we will be limited to just $\mathfrak{su}(N)$ gauge algebras over each curve. In the case of 6D SCFTs,
it is typically possible to obtain a rich class of possible sequences of gauge group factors, because 6D anomaly cancellation can be satisfied by introducing an appropriate number of additional flavor symmetry factors. This in turn leads to a notion of a ``ramp'' in the increase in the ranks \cite{DelZotto:2014hpa} (see also \cite{Hanany:1997gh}). For an affine quiver, this is much more delicate, since all of these cases can be viewed as a degeneration of an elliptic curve. For example, in the case of an affine $\widehat{A}_k$ base (i.e. the $I_{k}$ Kodaira type), anomaly cancellation tells us that all of the gauge algebra factors are the same $\mathfrak{su}(N)$. By a similar token, 6D anomaly cancellation tells us that the gauge algebra of any of these cases is $\mathfrak{su}(N d_i)$, where $d_i$ is the Dynkin label of the node in the affine graph, and $N \geq 1$ is an overall integer.\footnote{This same observation has already been made in the context of 4D superconformal $\mathcal{N} = 2$ quiver gauge theories \cite{Nekrasov:2012xe}. Indeed, in this special case the condition of vanishing beta functions is identical to the condition that 6D anomalies cancel.} In F-theory language, we have fiber enhancements $I_{N d_i}$ over each node. Indeed, at a formal level we can think of anomaly cancellation being satisfied by introducing $\mathfrak{su}(1)$ gauge algebras.

\subsubsection{The $1,2,...,2,1$ Base}

Compared with the case of affine ADE bases, there are comparatively more options available for fiber enhancements of the base $1,2,...,2,1$. In some sense, this is because these bases do not directly arise from the degeneration of a compact elliptic curve, but are better viewed as the degeneration of a cylinder.

With this in mind, we now explain how fiber enhancements work for this choice of base. For a large number of $-2$ curves, the allowed enhancements take a rather simple form, whereas there are outlier LSTs for smaller numbers of $-2$ curves.  In particular, when there are more than five $-2$ curves, the $-2$ curves necessarily hold $\mf{su}(N_i)$ gauge algebras:
$$
\overset{\mf{g}_L}1 \,\, \overset{\mf{su}(N_1)}2 \,\,\overset{\mf{su}(N_2)}2 ... \,\,\overset{\mf{su}(N_{k-1})}2 \,\,\overset{\mf{su}(N_{k})}2 \,\, \overset{\mf{g}_R}1
$$
The $N_i$ are subject to the convexity conditions $2 N_i \geq N_{i-1} + N_{i+1}$, with the understanding that $N_i =1 $ for a curve without a gauge algebra.

The $-1$ curves in this configuration may hold either $\mf{sp}(M)$ or $\mf{su}(M)$ gauge algebra.  If the leftmost $-1$ curve holds $\mf{sp}(M)$, anomaly cancellation imposes the additional conditions $2 N_1 \geq 2 M + N_2$, $2 M+ 8 \geq N_1$.  If the leftmost $-1$ curve holds $\mf{su}(M)$, anomaly cancellation imposes $2 N_1 \geq M + N_2$, $M+8 + \delta_{M,3} + \delta_{M,6} \geq N_1$.  Finally, the $-1$ curve may be empty provided $N_1 \leq 9$.  The story is mirrored for the rightmost $-1$ curve at the other end of the chain.

When there are exactly five $-2$ curves, we have two additional configurations:
$$
1 \,\, 2 \,\, \overset{\mf{su}(2)}2  \,\, \overset{\mf{so}(7)}2  \,\, \overset{\mf{su}(2)}2  \,\, 2 \,\, 1
$$
and
$$
1 \,\, 2 \,\, \overset{\mf{su}(2)}2  \,\, \overset{\mf{g}_2}2  \,\, \overset{\mf{su}(2)}2  \,\, 2 \,\, 1
$$

When there are four $-2$ curves, we similarly have
$$
1 \,\,  \overset{\mf{su}(2)}2  \,\, \overset{\mf{so}(7)}2  \,\, \overset{\mf{su}(2)}2  \,\, 2 \,\, 1
$$
and
$$
1 \,\, \overset{\mf{su}(2)}2  \,\, \overset{\mf{g}_2}2  \,\, \overset{\mf{su}(2)}2  \,\, 2 \,\, 1
$$

When there are three $-2$ curves, we have several new configurations:
$$
1 \,\,  \overset{\mf{su}(2)}2  \,\, \overset{\mf{so}(7)}2  \,\, \overset{\mf{su}(2)}2 \,\, 1
$$
$$
1 \,\, \overset{\mf{su}(2)}2  \,\, \overset{\mf{g}_2}2  \,\, \overset{\mf{su}(2)}2   \,\, 1
$$
$$
\overset{\mf{g}_L}1 \,\,   \overset{\mf{so}(7)}2  \,\, \overset{\mf{su}(2)}2  \,\, 2 \,\, 1
$$
and
$$
\overset{\mf{g}_L}1 \,\,  \overset{\mf{g}_2}2  \,\, \overset{\mf{su}(2)}2  \,\, 2 \,\, 1
$$
with $\mf{g}_L = \mf{sp}(M), M \leq 3$ in each of the last two cases.

When there are two $-2$ curves, we have:
$$
\overset{\mf{g}_L}1 \,\,   \overset{\mf{so}(7)}2  \,\, \overset{\mf{su}(2)}2   \,\, 1
$$
and
$$
\overset{\mf{g}_L}1 \,\,  \overset{\mf{g}_2}2  \,\, \overset{\mf{su}(2)}2  \,\, 1
$$
with $\mf{g}_L = \mf{sp}(M), M \leq 3$ in each of the last two cases.

When there is only a single $-2$ curve, there are even more possibilities:
$$
\overset{\mf{g}_L}1 \,\,   \overset{\mf{so}(8)}2  \,\,  \overset{\mf{g}_R}1
$$
with $\mf{g}_L = \mf{sp}(M_L), M_L \leq 2$, $\mf{g}_R = \mf{sp}(M_R), M_R \leq 2$.
$$
\overset{\mf{g}_L}1 \,\,   \overset{\mf{so}(7)}2  \,\,   \overset{\mf{g}_R}1
$$
with $\mf{g}_L = \mf{sp}(M_L)$, $\mf{g}_R = \mf{sp}(M_R)$, $M_L + M_R \leq 4$ or $M_L =4$, $M_R = 1$.
$$
\overset{\mf{g}_L}1 \,\,  \overset{\mf{g}_2}2  \,\, \overset{\mf{g}_R}1
$$
with $\mf{g}_L = \mf{sp}(M_L)$, $\mf{g}_R = \mf{sp}(M_R)$, $M_L + M_R \leq 4$.
$$
\overset{\mf{g}_L}1 \,\,  \overset{\mf{su}(2)}2  \,\, 1
$$
with $\mf{g}_L = \mf{g}_2$ or $\mf{so}(7)$.

\subsection{Loop-like Bases}

Finally, let us turn to the case of fiber enhancements for the loop-like bases. We have already discussed the case of an affine $\widehat{A}_k$
base of $-2$ curves in which we the allowed fiber enhancements are just a uniform $I_N$ fiber. Otherwise, we induce some blowups. Now, if we
allow for blowups, we can reach more general loop-like configurations. However, as we have already discussed near line (\ref{LoopConfMat}), all of these cases consist of a single type of base node suspended between minimal links. This is a consequence of the fact that in a general 6D SCFT, there are nested containment relations on the minimal fiber types \cite{Heckman:2015bfa}:
\begin{equation}
\mathfrak{g}^{min}_{1} \subseteq ... \subseteq \mathfrak{g}^{min}_{m} \supseteq ... \supseteq \mathfrak{g}^{min}_{k}.
\end{equation}
However, in a 6D LST, we must \textit{also} demand periodicity of the full
configuration. This forces a uniform fiber enhancement on each such node.

As a consequence, we can summarize all of these cases by keeping implicit the blowups associated with conformal matter. We have:
\begin{equation}
//\overset{\mathfrak{g}}2 , ..., \overset{\mathfrak{g}}2//
\end{equation}
where we allow for a general fiber enhancement to an ADE type simple Lie algebra $\mathfrak{g}$ over each of the $-2$ curves.
For all cases other than the $\mathfrak{su}(N)$ gauge algebras, this in turn requires further blowups in the base, i.e. we have a configuration with
conformal matter in the sense of references \cite{DelZotto:2014hpa, Heckman:2014qba}. So in other words, all of these loop-like configurations are summarized by stating the number of $-2$ curves, and the choice of fiber type over any of the $-2$ curves.

\section{Embeddings and Endpoints \label{sec:EMBED}}

In the previous sections we presented a general classification of LSTs in F-theory. One of the crucial ingredients we have
used is that decompactifying any curve must return us to a collection of (possibly disconnected) SCFTs.
Turning the question around, it is natural to ask whether \textit{all} SCFTs embed in LSTs.

In this section we show that this is indeed the case. Moreover, there can often be more than one way
to complete an SCFT to an LST. To demonstrate such an embedding, we will need to show that there exists a deformation of a given LST F-theory background which takes us to the requisite SCFT. This can involve both K\"ahler deformations, i.e. motion onto a partial tensor branch, and may
also include complex structure deformations, i.e. a Higgsing operation.

With this in mind, we first demonstrate that all of the bases for 6D SCFTs embed in an LST base. A suitable tensor branch flow then takes us from the LST base back to the 6D SCFT base. Then, we proceed to show that the available fiber decorations for LSTs can be Higgsed down to the fiber decorations for an SCFT. The latter issue is somewhat non-trivial since the fiber decorations of an ADE-type base is comparatively less constrained when compared with their affine counterparts.

\subsection{Embedding the Bases}

We now show that all bases for 6D SCFTs embed in LST bases. To demonstrate that such an embedding is possible, it is convenient to use the terminology of ``endpoints'' for SCFTs introduced in reference \cite{Heckman:2013pva}, which we can also extend to the case of LSTs. Given a collection of curves for an SCFT base, we can consider blowdowns of all of the $-1$ curves of the configuration. Doing so, we shift the self-intersection of all curves touching this $-1$ curve according to the rule $x \rightarrow (x-1)$ for a curve of self-intersection $-x$. In the case where a $-1$ curve is interposed in between two curves, we have $x,1,y \rightarrow (x-1) , (y-1)$. After this first stage of blowdowns, we can then sometimes generate \textit{new} $-1$ curves. Iteratively blowing down all such $-1$ curves, we eventually reach a configuration of curves which we shall refer to as an ``endpoint.'' The set of all endpoints has been classified in reference \cite{Heckman:2013pva}, and they split up according to four general types:
\begin{align}
\text{Trivial Endpoint} &  \text{: }1\rightarrow \mathbb{C}^2 \\
\text{A-type Endpoint} &  \text{: }x_{1}...x_{k}\\
\text{D-type Endpoint} &  \text{: }2\overset{2}{x_{1}}...x_{k}\\
\text{E-type Endpoint} &  \text{: }22\overset{2}{2}22\text{, \ \ }%
22\overset{2}{2}222\text{, \ \ }22\overset{2}{2}2222
\end{align}
In fact, starting from such an endpoint we can generate all possible bases of 6D SCFTs by further blowups. Sometimes such blowups are required
to define an elliptic Calabi-Yau, while some can be added even when an elliptic fibration already exists. By a similar token, we can also take a fixed base, and then decorate by appropriate fibers.

Now, a central feature of this procedure is that the resulting
adjacency matrix retains the important property that it is positive
definite. Similarly, if we instead have a positive semidefinite adjacency
matrix, the resulting matrix will retain this property under further
blowups (or blowdowns) of the base.

To demonstrate that we can always embed an SCFT in an LST, it will therefore suffice to show that there is \textit{some} way to add additional curves to an SCFT endpoint such that the resulting LST defines a base. To illustrate the idea, suppose we have a 6D SCFT with a trivial endpoint. Then, before the very last stage of blowdowns, we have a single $-1$ curve, which we shall call $\Sigma$. If we return to the original SCFT, this curve will also be present, but its self-intersection will be different. Hence, to get an LST we can simply attach one more $-1$ curve to $\Sigma$. For example, the configuration $1,1$ defines the base of an LST. Supplementing the fibers can always be done to realize the case of fiber decorations.

Consider next the case of A-type endpoints. Here, we can always attach a suitable number of ``tails'' of the form $1,2...,2$ to each curve such that blowing down these instanton links leads us to $k$ curves of self-intersection $-2$, i.e. $2,...,2$. Attaching an additional $-1$ curve to the left and to the right, we get a base of the form $1,2,...,2,1$. We therefore conclude that adding such tails again allows an embedding in an LST.

By a similar token, we can append such tails to a D-type endpoint until we reach the Dynkin diagram with just $-2$ curves. Adding one more $-1$ curve to this configuration:
\begin{equation}
\text{D-type Endpoint} = 2\overset{2}{2}...2,1
\end{equation}
leads to a blowdown eventually to the configuration $2,1,2$, which in turn blows down to $1,1$. So again, we conclude that the adjacency matrix is positive semidefinite and we have arrived at an LST.

This leaves us with the E-type endpoints. In these cases, we just have a configuration of $-2$ curves, and the possible blowups are severely limited \cite{Heckman:2013pva}. For example, there are no blowups of the $E_8$ Dynkin diagram. It therefore suffices to add one additional $-2$ curve to this configuration to reach its affine extension. Similar considerations also apply for the $E_6$ and $E_7$ configurations when no additional blowups are present, i.e. we simply proceed from the Dynkin diagram to its affine extension.

To round out the analysis, we need to demonstrate that if we perform any blowups of an $E_6$ or $E_7$ endpoint, we can again attach a $-2$ curve at the same location, without inducing any further blowups.\footnote{Note that attaching a $-1$ curve will not work since we would then blowdown to a configuration where the adjacency matrix is not positive semidefinite -- a contradiction.} That this is indeed the case is conveniently summarized by simply writing down the possible blowups. For $E_6$, there are two other consistent bases, and for $E_7$ there is one. In both cases, we can indeed still add our $-2$ curve without inducing extraneous blowups:
\begin{align}
2321\overset{2}{\overset{3}{\overset{2}{\overset{1}{8}}}}1232 &
\rightarrow231\overset{2}{\overset{3}{\overset{1}{5}}}132\rightarrow
22\overset{2}{\overset{2}{2}}22\\
2231\overset{3}{\overset{1}{5}}1322 &  \rightarrow222\overset{2}{2}222.
\end{align}

Summarizing, we have just demonstrated that \textit{all} bases for
6D SCFTs embed in an LST base.

\subparagraph{LST Endpoints}

As a brief aside, one of the interesting features of this argument is that we have implicitly relied on the notion of an LST endpoint. Given that
we have already classified all such bases, we can also ask about the possible endpoints for LSTs. Compared with the
case of 6D SCFTs, the number of distinct endpoints are comparatively small. Roughly speaking, this is because of the positive semidefinite
condition for our adjacency matrix, which in turn means that many configurations will blow down to a single curve of self-intersection zero (as in the $1,2,...,2,1$ configurations).

Combining the tensor-decoupling criterion with the demand that we have a positive semidefinite adjacency matrix means that the total number of endpoints are given by the Kodaira type intersections of $-2$ curves, an elliptic curve, as well as a single $\mathbb{P}^1$ of self-intersection zero, which we denote by $P$. In the latter two cases, we note that we only obtain a 6D LST by having a non-trivial elliptic fibration.  Thus, we find the following list of LST endpoints:\footnote{Here we neglect the possibility of torsion in the normal direction to the compact curves of the base.}
\begin{equation}
\text{LST Endpoints:}\,\,\,P,I_{n},II,III,IV,I^{\ast}_{n},II^{\ast},III^{\ast},IV^{\ast}%
\end{equation}
for $n \geq 0$.

\subsection{Embedding the Fibers}

Suppose next that we have supplemented a base by an additional curve. When we do this, additional non-trivial fibers are sometimes inevitable, and can in turn force additional structure on the elliptic fibers. To give a concrete example, consider the case of a 6D SCFT with base given by the $E_6$ Dynkin diagram of $-2$ curves. Fiber decorations for this model were studied in reference \cite{Heckman:2015bfa} where it was found that typically,
additional flavors can be added so that an appropriate convexity condition on the ranks is obeyed. To extend this to an LST base, we cannot add a $-1$ curve (as the blowdown is inconsistent). Rather, we must add an additional $-2$ curve to reach the affine $\widehat{E}_6$ Dynkin diagram of $-2$ curves. When we do this, we must remember that the elliptic fibration of the resulting F-theory model also becomes rigid. So in other words, the available elliptic fibrations are further constrained. It is at this stage that we must include the effects of Higgsing as well as tensor branch deformations to reach the original 6D SCFT. That this is always possible follows from the fact that all representations of the Lie algebra $\mathfrak{e}_6$ embed in representations of its affine extension $\widehat{\mathfrak{e}}_6$. Similar considerations apply for the fiber decorations of all of the E-type bases.

In the case of the A- and D-type bases, the analysis is comparatively simpler. The reason is that we can just add additional tails of the form $1,2,...,2$ with trivial fibers and leave the fibers above the curves in the SCFT base as they were.  One might worry here about the fact that curves of self-intersection $-1$ and $-2$ are not always allowed to have trivial fibers.  For instance, the $-1$ curve in the sequence $2,2,3,1,5$ necessarily has a type $II$ fiber.  However, such subtleties do not arise in this case: one can always add small instanton links of the form $1,2,...,2$ with trivial fibers to an A- or D-type base to get an LST.

Putting together our analysis of tensor branch flows and Higgs branch flows, we conclude that all 6D SCFTs embed in some LST. Indeed, it is also clear that there can sometimes be more than one such embedding.

\section{T-duality \label{sec:Tduality}}

In the previous sections we used the geometry of F-theory compactifications to
tightly constrain the structure of LSTs. In this section we turn the analysis
around and show how the physics of little string theories suggests non-trivial
geometric structures for elliptic Calabi-Yau threefolds in which the
non-compact base has a negative semidefinite intersection form.

In physical terms, one of the key features of LSTs is that the description as
a local quantum field theory must break down near the string scale. A sharp
way to probe this structure is by compactifying on a circle. Recall that in
T-duality, the theory compactified on a small $S^{1}$ of radius $R$ is dual to
another string theory compactified on an $S^{1}$ of radius $\widetilde{R}%
\sim \alpha^{\prime}_{eff}/ R$. Based on this, it is natural to expect that
all LSTs have a similar T-duality.

This expectation suggests a non-trivial constraint on the geometry of an F-theory
realization of an LST. Recall that F-theory compactified on an elliptically
fibered Calabi-Yau threefold $X\rightarrow B$ leads, upon further
compactification on an $S^{1}$, to M-theory compactified on the same
Calabi-Yau threefold. In the M-theory description, the K\"{a}hler class
becomes a dynamical modulus (which is taken to zero size to reach the F-theory
limit). On the other hand, T-duality tells us that if we take this circle to
be very small, we should expect to obtain another LST, this time compactified
on a large radius circle. For this to be so, we must have available to us more
than one way to reach an F-theory background from a given compactification of
M-theory on $X$. In other words, physical considerations suggest the existence
of \textit{another} elliptic fibration for our Calabi-Yau threefold
$X\rightarrow\widetilde{B}$, and the lift from M-theory to F-theory involves
collapsing the K\"{a}hler class of this other elliptic fiber to zero size.

Our plan in this section will be to give further evidence that T-duality is
realized in F-theory constructions of LSTs through the presence of a double
elliptic fibration. We shall, however, mainly focus on particular examples of
how T-duality is realized geometrically. After this, we give a
sketch for how we expect this correspondence to work in general,
leaving a complete proof to future work.

\subsection{Examples}

As a first example consider the T-duality between the LST of $k \geq 2$
NS5-branes in IIA (the $\mathcal{N} = (2,0)$ A-type LSTs), and that of $k$ NS5-branes in
IIB\ string theory (the $\mathcal{N} = (1,1)$ A-type LSTs). The IIA\ realization just
follows from a base with $-2$ curves arranged in a
loop, i.e. as the type $I_{k-1}$ degeneration of an elliptic curve. The F-theory
elliptic curve is then a smooth $T^{2}$, i.e. an $I_{0}$ fiber. Switching the
roles of these two curves, we get the type IIB\ $\mathcal{N}=(1,1)$ LST, i.e. we have $(k-1)$
D7-branes wrapped over a $T^{2}$. There is a clear extension of this case to
all of the ADE $\mathcal{N}=(2,0)$ LSTs in terms of the corresponding ADE 7-branes
wrapped over a $T^{2}$.

Another class of examples we have already encountered several times involves
the LSTs realized by M5-branes filling $\mathbb{R}^{5,1}$ and probing the
geometry $S_{\bot}^{1}\times\mathbb{C}^{2}/\Gamma$ for $\Gamma$ an
ADE\ subgroup of $SU(2)$. Compactifying on a further circle, we can shrink the
$S_{\bot}^{1}$ factor to reach IIA string theory. T-duality is then inherited from that of
the physical superstring theory.

In the F-theory realization of these systems, we have a base quiver with
conformal matter suspended in between the nodes:%
\begin{equation}
//\underset{k}{\underbrace{g\oplus g\oplus...\oplus g\oplus g}}//.
\end{equation}
As we have already remarked in the classification of such structures, all
fiber types and conformal matter are necessarily minimal; no deviations from
this rigid structure are possible. This actually means that we can readily
identify the other elliptic fiber of this model: It is given by a suitable
multiple of an $I_{kd_{i}}$ fiber, where the $d_{i}$ denote the Dynkin labels
of the affine extension of the $g$-type Lie algebra. Another consequence of
this analysis is that sometimes, the absence of other fiber
decorations for say the E-type affine bases means we cannot arbitrarily
combine these two structures. Rather, since the only fiber enhancements over
the E-type bases are $I_{k}$-type fibers (i.e. without inducing further
blowups), there is again a clear exchange between the roles of the two
(singular) elliptic curves.

Quite strikingly, this also entails the existence of several infinite classes
of models which are self-T-dual upon toroidal reduction. These are models
which have a double elliptic structure consisting of two copies of the \textit{same}
Kodaira fiber. For example, models which have an $I_{N}$ Kodaira
base, with gauge groups $\mathfrak{su}(N)$ on each $-2$ curve and bifundamentals at
the intersections. As a more exotic example consider the blown up type
$IV$ degeneration: $3\overset{3}{1}3$. On each $-3$ curve in this configuration there is an
$\mathfrak{su}(3)$ gauge sector which arises precisely from a type $IV$
fiber. Clearly, contracting the -1 curve, switching fiber and base, and then
blowing up we obtain back the same model.\footnote{See \cite{MPT} for a
further analysis of this case.}

As a final class of examples, consider the F-theory models with base:%
\begin{equation}
[E_8] \underset{k}{\underbrace{1,2,...,2,1}} [E_8].
\label{eq:ex-1221}
\end{equation}
This is realized in heterotic M-theory by a collection of $k$ M5-branes in
between the two heterotic walls. compactifying on a further circle and
activating appropriate Wilson lines for the background $E_{8}\times E_{8}$
flavor symmetry, we reach --via T-duality of the physical superstring-- the
case of $k$\ NS5-branes of the $Spin(32)/%
\mathbb{Z}
_{2}$ heterotic string theory. This system can be analyzed in perturbative string theory, and
the T-dual LST is therefore realized by an $\mathfrak{sp}(k)$ gauge theory with $32$ half hypermultiplets
in the fundamental representation, and a single hypermultiplet in the two-index anti-symmetric representation.

Demonstrating the presence of the extra elliptic fibration for the F-theory
model is somewhat more subtle in this case, but we can see it as descending
from a $%
\mathbb{Z}
_{2}$ quotient of the $I_{2k}$-type Kodaira configuration of $-2$ curves:%
\begin{equation}
//\underset{2k}{\underbrace{2,2,...,2,2}}//\overset{\mathbb{Z}_{2}%
}{\rightarrow}\underset{k}{\underbrace{1,2,...,2,1}}.
\end{equation}
In Appendix \ref{appx:Texamp} we present an explicit analysis of this case of the base $1,1$, and also explain its extension to the configuration
$1,2,...,2,1$. The corresponding F-theory model is realized by a base given by $P$, a single $\mathbb{P}^1$ with self-intersection zero,
with a fiber enhancement $I_{2k}^{ns}$, i.e. we get an $\mathfrak{sp}(k)$ 7-brane wrapped over $P$. We have already classified
the matter enhancements for this case in section \ref{sec:FIBERS}, and indeed, we find agreement with the purely heterotic analysis. Note that the
$\mathfrak{sp}$-type algebra originates from the $\mathbb{Z}_2$ quotient of an $A_{2k}$ algebra via its outer automorphism.

In fact, a similar observation allows us to extend this to some of the models
in which we have a $\mathbb{P}^{1}$ of self-intersection zero. For example,
under a further $\mathbb{Z}_{2}$ quotient, we can reach some of the gauge
theories already encountered via more direct methods:%
\begin{equation}
\lbrack\mathfrak{su}_{8}]\overset{\mathfrak{su}_{N}}{1},\overset{\mathfrak{su}%
_{N}}{1}[\mathfrak{su}_{8}]\overset{\mathbb{Z}_{2}}{\rightarrow}%
[\mathfrak{su}_{16}]\overset{\mathfrak{su}_{N}}{0}[\mathfrak{su}_{2}],
\end{equation}
where the $\mathfrak{su}_{16}$ flavor symmetry acts on the sixteen
hypermultiplets in the fundamental representation, and the $\mathfrak{su}_{2}$
flavor symmetry acts on the hypermultiplets in the two index anti-symmetric
representation. From this perspective, we can still recognize the quotient of
an additional elliptic fiber. Schematically, we have:
\begin{equation}
// 2 , 2 // \overset{\mathbb{Z}_2}{\rightarrow} 1,1 \overset{\mathbb{Z}_2}{\rightarrow} 0.
\end{equation}

\subsection{Examples Involving Curves with Torsion Normal Bundle}

Recall that in section \ref{sec:EXAMPLES} we found that the $\mathcal{N}=(1,1)$ LSTs admitted an ABCDEFG classification
according to their corresponding affine Lie algebras. In the F-theory realization of these models, we also saw that the base contained a genus $1$ curve with a torsion normal
bundle.  We encountered torsion normal bundles again in section \ref{sec:RANKZERO} in our discussion
of rank zero LSTs.  We will now explain these examples in more detail.

By construction, examples in this class admit two genus one fibrations:
$(T^2_F\times S)/\mathbb{Z}_m \to (T^2_F\times C)/\mathbb{Z}_m$,
and $(T^2_F\times S)/\mathbb{Z}_m \to S/\mathbb{Z}_m$.
The first fibration has a section with monodromy over the central fiber
of $\pi: (T^2_F\times C)/\mathbb{Z}_m \to C$.  The second fibration does
not have a section
(i.e., it is not an elliptic fibration,
in the terminology of \cite{Braun:2014oya}),
and we discuss its structure here.

Recall that for a genus one fibration $X\to B$ without a section, there is an
associated ``Jacobian fibration'' $J(X/B)\to B$ which has a section, and which has
precisely the same $\tau$ function describing the
fibers.\footnote{A more common notation is $J(X)$, but since we are
studying Calabi--Yau threefolds with more than one genus one fibration,
we indicate the fibration for emphasis.}  As explained in
\cite{deBoer:2001px,Braun:2014oya,multisections}, the set of $X$'s which share
a common Jacobian fibration (and are equipped with an action by the Jacobian
fibration, as stressed in \cite{Cvetic:2015moa}) forms a group\footnote{This group has been incorrectly
called the Tate--Shafarevich group in the literature \cite{Braun:2014oya,multisections},
for which the fourth author apologizes.  The group contains the Tate--Shafarevich
group \cite{MR1242006}, but it can also contain fibrations with isolated multiple
fibers \cite{MR1401771}, a fact which  arose in the proof of the
finiteness theorem for elliptic Calabi--Yau threefolds \cite{alg-geom/9305003,MR1272978}.
The examples presented here have such isolated multiple fibers, and so do not
belong to the Tate--Shafarevich group but rather to the larger group of Calabi--Yau
threefolds sharing a common Jacobian fibration.} which should be identified
with the group of connected components of the gauge group in F-theory.  That is,
compactifying the F-theory model on a circle, there is a discrete choice of one of
these  $X$'s to serve as the compactification space for M-theory, which is the hallmark
of a discrete gauge choice.

In our examples, the action of $\mathbb{Z}_m$ on $S$ has fixed points, leading
to $A_{m-1}$ singularities on the quotient $S/\mathbb{Z}_m$.  (More
generally, there can be $A_{\ell-1}$ singularities for any $\ell$ dividing
$m$, due to fixed points of subgroups.)  Over an $A_{\ell-1}$
 point, we have
taken the quotient of the fiber by a translation of order $\ell$,
so that fiber has multiplicity $\ell$.  Notice that even though the
fiber is multiple, the total space is smooth.

This is precisely the situation analyzed by Mark Gross in \cite{MR1401771},
who showed that the Jacobian fibration is fibered over the same base,
still with $A_{\ell-1}$ singularities.  But once the Jacobian fibration has
been taken, it is possible to resolve those singularities of the base
(which corresponds
physically to giving an expectation value to scalars in the corresponding tensor
multiplets).  We thus find that these theories are part of a larger family
of LSTs, but at special values of the tensor moduli in the larger family,
a finite gauge group appears, leading to additional
5D vacua (i.e., the fibrations without a section)
corresponding to distinct sectors of Wilson line expectation values.

Let us illustrate this with two concrete examples drawn from
Appendix~\ref{app:F1-1}, where we work out the $\mathcal{N}=(1,1)$ theories
of BCFG type in detail.  As a first example, consider  a partially blown
down graph of type $\widehat{E}_8$, illustrated here (and in
Figure~\ref{figure-F4}):

\begin{center}
\includegraphics{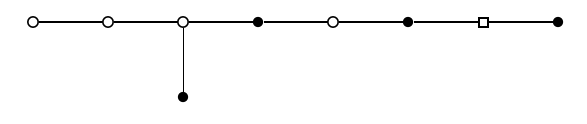}
\end{center}

\noindent
where we blow down each solid circle to an $A_1$ singularity.  As shown
in Appendix~\ref{app:F1-1}, there is a genus $1$ fibration $X$ over this base $B$ with a
fiber of multiplicity $2$ at each $A_1$ singularity.  The fiber never
degenerates over this locus.

The Jacobian fibration
$J(X/B)$ is simply a Weierstrass fibration over $B$ whose fibers do not
degenerate.  There is no obstruction to resolving the singularities
of $B$, moving out into the rest of the moduli space.  In fact, this is
part of the moduli space of the
$\mathcal{N}=(2,0)$ theory of type $E_8$, and we claim that when the
tensors are tuned to blow down precisely the curves corresponding
to solid nodes,  a $\mathbb{Z}_2$ gauge symmetry
appears in the theory.
That gauge symmetry is necessary to explain the additional 5D
models which appear when the $A_1$'s are blown down and a twist of the
Jacobian fibration is possible.  Presumably, moving away from this locus
amounts to Higgsing the $\mathbb{Z}_2$ gauge symmetry.
Incidentally, this example is the T-dual
of the $\mathcal{N}=(1,1)$ model with group $F_4$.

As a second example (which is the T-dual of the $\mathcal{N}=(1,1)$ model with group $G_2$),
consider  another partially blown
down graph of type $\widehat{E}_8$, illustrated here (and in
Figure~\ref{figure-G2}):

\begin{center}
\includegraphics{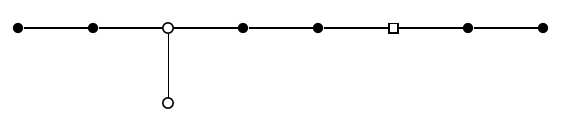}
\end{center}

\noindent
where this time, we blow down each lined pair of solid circles to an $A_2$ singularity.
There is a fibration with multiple fibers of multiplicity $3$ over
each of those points.  The Jacobian fibration is again part of the $\mathcal{N}=(2,0)$
theory, with a different tuning of the moduli.  With this tuning, we
find a $\mathbb{Z}_3$ gauge symmetry.

A related example was given in section \ref{sec:RANKZERO} using the action
of $\mathbb{Z}_6$ on a nonsingular elliptic curve with $j=0$.  The quotient
$S/\mathbb{Z}_6$ has central fiber which is a rational curve passing through
surface singularities of types $A_1$, $A_2$, and $A_5$ with fibers of
multiplicity $2$, $3$, and $6$ over those points.  Resolving the singularities
leads to another affine $\widehat{E}_8$ diagram:

\begin{center}
\includegraphics{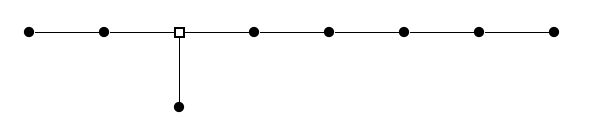}
\end{center}

\noindent
where again the solid circles represent the curves being blown down.
Tuning the moduli of the $\mathcal{N}=(2,0)$ $E_8$ theory to blow down those curves
leads to a locus in moduli with a $\mathbb{Z}_6$ gauge symmetry.
Notice that this is the intersection of the previous two loci with
$\mathbb{Z}_2$ and $\mathbb{Z}_3$ gauge symmetry.

\subsection{Towards T-Duality in the More General Case}

In the examples from the previous two subsections,  we saw that
the expected appearance of T-duality for an LST motivates the search for a double elliptic fibration structure
in such F-theory models.
When the base has a fibration by curves of genus $1$, the origin of this
second fibration is clear.  When the fibration on the base is by curves
of genus $0$, however, the T-duality is not as readily manifest.

Example~\eqref{eq:ex-1221} does have manifest T-duality,
 as further analyzed in
Appendix~\ref{appx:Texamp}.
In this case, the Weierstrass model $y^2-F(x,\psi,[s,t])$
can be regarded as a double cover of a $\mathbb{P}^1$-bundle over
the base $B$, where $x$ denotes the coordinate on the $\mathbb{P}^1$.
Since the double cover is branched at $4$ points along each fiber
 $\mathbb{P}^1_F$ of this fibration,
the total space gets a fibration by curves of genus $1$.

Now the base $B$ of that F-theory model has a fibration
 $\pi:B \to C$ of its own whose general fiber is another $\mathbb{P}^1$
which we shall call $\mathbb{P}^1_B$.
In Appendix~\ref{appx:Texamp}, after making a birational modification of
the base (i.e., blowing it up and down),
we find that the double cover is branched along $4$ points of each fiber
$\mathbb{P}^1_B$ as well, and this implies that there is a {\em second}\/
genus $1$ fibration on the total space.
Note that this echoes the discovery made in \cite{Aspinwall:1997ye} that
heterotic T-duality, when viewed from the perspective of F-theory, exchanges
the roles of base and fiber in the heterotic weak coupling limit.

It is therefore natural to seek out a more general geometric
exchange symmetry in LSTs with T-duality. We leave a more complete investigation of
this possibility to future work.

\section{Outliers and Non-Geometric Phases \label{sec:OUTLIERS}}

Much as in the case of the classification of SCFTs achieved in reference \cite{Heckman:2015bfa}, we view the F-theory realization of LSTs as providing a systematic approach to the construction of such models. In some cases, however, we indeed find a few small gaps between what is expected based on field theory considerations, and what can be obtained in geometric phases of F-theory.

Our plan in this short section will be to proceed mainly by effective field theory considerations to give a list of such outlying behavior, both for 6D SCFTs and LSTs. We hasten to add that some of these putative theories may end up being inconsistent due to the lack of an embedding in an F-theory (or other string) construction.\footnote{For a recent example of a seemingly consistent 6D SCFT which is actually inconsistent, see e.g. Appendix A of reference \cite{Ohmori:2015pia}.} In some cases, however, this also points to a few additional novel possibly non-geometric structures. Though we shall comment on possible ways to realize these models in F-theory, we leave a more complete analysis to future work.

\subsection{Candidate LSTs and SCFTs}

From a bottom up perspective, the primary constraints on the construction of consistent LSTs are the existence of a lattice of string charges with a negative semidefinite Dirac pairing, and possible gauge groups ``decorating'' the associated tensor multiplets.

In some cases, there is a clear indication from F-theory that certain bottom up considerations are too weak; For example, the phenomenon of a $-n$ curve for $ n \geq 3$ always implies the existence of a non-trivial gauge group factor, a condition which is not obvious from any anomaly cancellation condition.

There are, however, three intermediate cases suggested by field theory which also have a potential realization in string theory.  The first case deals with the gauge-gravitational anomaly cancellation condition (see (\ref{firstanomaly})) imposed in all geometrically realized 6D SCFTs and LSTs. This condition has no field theoretic analog in flat space as there are no a priori restrictions on one-loop contribution to mixed gauge-gravitational anomaly. Non-vanishing of total mixed gauge-gravitational anomaly implies that the theory is inconsistent when put on a fixed curved spacetime background. It would clearly be troublesome if 6D SCFTs were anomalous in this way. Fortunately, it is always possible to cancel the one-loop contribution to this anomaly against a Green-Schwarz contribution, as was demonstrated in \cite{Ohmori:2014kda}.  This argument does not apply to 6D LSTs, and indeed it can be checked explicitly in many examples of 6D LSTs that there is no way to cancel the one-loop mixed gauge-gravitational anomaly using the Green-Schwarz mechanism. This means that 6D LSTs cannot always be put on a fixed curved spacetime.

One class of such models would arise on:
\begin{align}
12...21
\end{align}
where we decorate all the tensors with $\mathfrak{su}$-type gauge groups along with anti-symmetric matter for the last $\mathfrak{su}$ gauge group and symmetric matter for the first $\mathfrak{su}$ gauge group. With the presence of this symmetric representation, (\ref{firstanomaly}) is violated. Nonetheless, this model can be constructed by putting type IIA on an interval $S^1/\mathbb{Z}_2$ with an $O8^{-}$ orientifold plane on each fixed point, D6s stretched along the interval, NS5s embedded in the D6s at various points along the interval, and two of the NS5s stuck respectively at the two fixed points.

More generally, to construct new examples of 6D SCFTs and LSTs which violate the 1-loop mixed gauge-gravitational anomaly condition (\ref{firstanomaly}) we can do the following: take any F-theory model having a $-1$ curve or $P$ base curve associated with $\mathfrak{su}$-type gauge group and at least one anti-symmetric hyper and 16 fundamental hypers not transforming under any other gauge group. Then, replace this set of hypermultiplets with a single hypermultiplet in the symmetric representation.  The resulting theory will satisfy gauge anomaly cancellation, but it will have a non-vanishing 1-loop gauge-gravitational anomaly.

The second case has to do with the condition of ``normal crossing'' which is present in all geometrically realized 6D SCFTs, and is only mildly violated in LSTs. For example, we have seen that an intersection with an order two tangency $4 || 1$ leads to a consistent LST base. In the bottom up perspective, we have a negative semidefinite Dirac pairing which has $2$ on the off-diagonal entries, and $-4$ and $-1$ on the diagonals. Generalizing, we can consider constructions such as:
\begin{align}
& 4||2...2\\
& 4||2...21\\
& 4||2...2||4
\end{align}
all of which have a negative definite Dirac pairing. Though we have not encountered these possibilities in our discussion of geometric phases of F-theory, the first two have been realized in IIA string theory via appropriate suspended brane configurations (see e.g. \cite{Hanany:1997gh}), at least when there are non-trivial gauge group factors over the associated tensor multiplets. For example, we can have $\mathfrak{su}$-type on the $-2$ charge tensors, and $\mathfrak{so}$-type gauge groups on the $-4$ tensors, with $\mathfrak{sp}$-type gauge groups on the $-1$ tensors. The crucial ingredient appearing in these suspended brane models is an $O8^{+}$-orientifold, rather than an $O8^{-}$-orientifold. The T-dual description in IIB string theory involves $O7^{+}$-orientifold planes, a case which leads to some non-geometric behavior, a point we shall return to later. The last example $4||2...2||4$ resists an embedding in IIA string theory, since it would appear to involve two $O8^{+}$-orientifold planes. Indeed, if we attempt to decorate these tensor multiplets by $\mathfrak{so}$-type algebras on the $-4$ tensors, and $\mathfrak{su}$-type algebras on the $-2$ tensors, we cannot cancel gauge theoretic anomalies.

Assuming that configurations such as $4 || 2$ can indeed occur in the construction of LSTs, it is natural to ask how many additional models can be obtained, at least from a bottom up perspective. Pairing each tensor multiplet with a simple gauge algebra, these are as follows:
\begin{gather}
\overset{\mathfrak{so}(M)}4 \,\, || \,\, \overset{\mathfrak{su}(N_1)}2 \,\, \overset{\mathfrak{su}(N_2)}2 \,\,... \,\, \overset{\mathfrak{su}(N_k)}2 \,\, \overset{\mathfrak{sp}(N_R)}1 \\
\overset{\mathfrak{sp}(N_1)}1\,\, \overset{\mathfrak{sp}(N_T)}{\overset{1}{\overset{\mathfrak{so}(M_1)}{4}}}  \,\, ... \,\, \overset{\mathfrak{sp}(N_k)}1 \,\,\overset{\mathfrak{so}(M_k)}4 \,\, || \,\,\overset{\mathfrak{su}(N_R)}2 \\
\overset{\mathfrak{su}(N')}2 \,\, \overset{\mathfrak{sp}(N_1)}1 \overset{\mathfrak{so}(M_1)}4 \,\, ... \,\,\overset{\mathfrak{so}(M_k)}4 \,\, || \,\,\overset{\mathfrak{su}(N_R)}2
\end{gather}
It should be noted that gauge and mixed anomalies strongly constrain the allowed gauge algebras in the above list.  In particular, the configuration $\overset{\mathfrak{so}(M)}4 || \overset{\mathfrak{su}(N)}2$ is constrained by mixed anomalies to have a bifundamental $(\bf{M, N})$, requiring
$N \leq M -8$, $2 N \geq M$.  These conditions lead to strong constraints on the rest of the gauge algebras in the aforementioned theories, as discussed in \cite{Bhardwaj:2015xxa}.

If one also relaxes the condition that there is a gauge group paired with each tensor, even more constructions are possible. In most of these cases, no known embedding in a string construction is available, so we suspect that at least some of these theories are actually inconsistent. Nevertheless, for the sake of completeness, we list them here:
\begin{gather}
2  \,\,||  \,\,4 \,\,1 \,\,4 \,\, ... \,\,4 \,\,1 \,\,4  || \,\, 2 \label{limit} \\
4 \,\, 1 \,\, 4 \,\, || \,\, 2 \,\, 2 \\
4 \,\, ||  \,\,2 \,\,2 \,\, ... \,\, 2 \,\, || \,\, 4\\
4 \,\, ||  \,\,2 \,\,2 \,\, ... \,\, 2 \,\, 1\\
4 \,\, ||  \,\,2 \,\,2 \,\, ... \,\, \overset{2}2 \,\, 2 \\
1\,\,\overset{1}{4}\,\,1\,\,... \,\,4 \,\, 1 \,\, 4 \,\, || \,\, 2\\
1\,\, 3 \,\,1\,\,... \,\,4 \,\, 1 \,\, 4 \,\, || \,\, 2\\
1\,\, 2\,\,3 \,\, 1\,\,4\,\,... \,\,4 \,\, 1 \,\, 4 \,\, || \,\, 2\\
1\,\, 2\,\,2\,\, 3 \,\, 1\,\,4\,\,... \,\,4 \,\, 1 \,\, 4 \,\, || \,\, 2\\
2 \,\, 1\,\,4\,\,... \,\,4 \,\, 1 \,\, 4 \,\, || \,\, 2.
\end{gather}

The third case can be viewed as a limit of (\ref{limit})
\begin{gather}
2  \,\,||  \,\, 2
\end{gather}
with the following gauge algebras
$$
\,\, \underset{[Sp(1)]}{\overset{\mf{so}(7)}2} \,||\, \overset{\mf{su}(4)}2
$$
$$
\,\, \underset{[Sp(2)]}{\overset{\mf{so}(8)}2} \,||\, \overset{\mf{su}(4)}2
$$
$$
\,\, {\overset{\mf{g}_2}2} \,||\, \overset{\mf{su}(4)}2
$$
$$
\,\,  \underset{[N_s=1]  }{\overset{\mf{so}(12)}2} \,||\, \overset{\mf{su}(6)}2
$$
$$
\,\,\underset{ [N_s=1/2]}{\overset{\mf{so}(13)}2} \,||\, \overset{\mf{su}(7)}2
$$
Their omission has to do with the fact that the $\mf{su}(N)$ for $N=4,6,7$ appearing above is constructed by an $I_N$ singularity on a $-2$ curve. But an $I_N$ singularity on a $-2$ curve cannot consistently intersect\footnote{This is because the order of vanishing of $f$ for an $I_N$ singularity is zero. If the $I_N$ singularity wraps a $-2$ curve $C$ in the base $B$, then the number of points on $C$ where $f$ vanishes is computed as $-4K_B\cdot C=0$ where $K_B$ is the canonical divisor on $B$. This is in contradiction with the fact that $C$ intersects another curve $D$ carrying an $I^*_M$ singularity on which $f$ vanishes at least to order two.} an $I^{*}_M$ type singularity required to construct $\mf{so}(7),\mf{so}(8),\mf{so}(12),\mf{so}(13),\mf{g}_2$ above.

It is worth mentioning that a violation of normal crossing and a violation of gauge-gravitational anomaly cancellation do not appear simultaneously in any of these examples.  The lists of LSTs and putative LSTs arising from these violations are small and tightly constrained.

This perspective on LSTs also points to the existence of some additional novel structures for 6D SCFTs.
Indeed, starting from an LST, we can consider a combination of tensor-decoupling and Higgsing to reach
some additional candidate SCFTs.

The theories of this type, which are consistent with anomaly cancellation and with no unpaired tensors, take the form:
\begin{gather}
\overset{\mathfrak{so}(M)}4 \,\, || \,\, \overset{\mathfrak{su}(N_1)}2 \,\, \overset{\mathfrak{su}(N_2)}2 \,\,... \,\, \overset{\mathfrak{su}(N_k)}2  \\
 \overset{\mathfrak{sp}(N')}1 \,\, \overset{\mathfrak{so}(M)}4 \,\, || \,\, \overset{\mathfrak{su}(N_1)}2 \,\, \overset{\mathfrak{su}(N_2)}2 \label{outoutlier} \\
 \overset{\mathfrak{so}(M_1)}4 \,\,  \overset{\mathfrak{sp}(N_1)}1 \,\,... \,\,\overset{\mathfrak{so}(M_k)}4 \,\, || \,\,\overset{\mathfrak{su}(N_R)}2 \\
 \overset{\mathfrak{sp}(N_1)}1 \overset{\mathfrak{so}(M_1)}4 \,\, ... \,\,\overset{\mathfrak{so}(M_k)}4 \,\, || \,\,\overset{\mathfrak{su}(N_R)}2
\end{gather}
Notice that (\ref{outoutlier}) does not admit a known type IIA construction, whereas the other three do. Another curious thing to notice is that this model does not admit an embedding in a putative LST. The obvious examples of attaching an $\mathfrak{so}$ group to the left or attaching an $\mathfrak{su}$ group to the right are pathological because they necessarily have non-vanishing quartic part of gauge anomaly. Although we have only proved that every 6D SCFT can be embedded in a 6D LST for models arising geometrically within F-theory, we expect this statement to be true in general. Hence, we suspect that the putative SCFT (\ref{outoutlier}) is inconsistent. Finally, concentrating only on positive-definiteness, we have the $4||2$ configurations:
\begin{gather}
 1 \,\, 4 \,\, || \,\, 2  \,\, 2\\
4 \,\, ||  \,\,2 \,\,2 \,\, ... \,\, 2 \\
1\,\, 4 \,\,1\,\,... \,\,4 \,\, 1 \,\, 4 \,\, || \,\, 2\\
4\,\, 1 \,\,4\,\,... \,\,4 \,\, 1 \,\, 4 \,\, || \,\, 2\\
3 \,\,1\,\,... \,\,4 \,\, 1 \,\, 4 \,\, || \,\, 2\\
 2\,\,3 \,\, 1\,\,4\,\,... \,\,4 \,\, 1 \,\, 4 \,\, || \,\, 2\\
 2\,\,2\,\, 3 \,\, 1\,\,4\,\,... \,\,4 \,\, 1 \,\, 4 \,\, || \,\, 2
\end{gather}

\subsection{Towards an Embedding in F-theory} \label{sec:towards-embedding}

In the above, we encountered some additional candidate SCFTs and LSTs which appear to be consistent with effective field theory considerations. Additionally, some of these models admit an embedding in type IIA suspended brane constructions.

It is therefore important to ask to what extent we should expect
F-theory to cover this and related examples. Though we leave a complete characterization
to future work, there are some general ingredients we can
already identify which point the way to incorporating these additional non-geometric
structures.

As we have already mentioned, one crucial ingredient in the IIA realization of the $4 || 2$ configuration is the appearance of an $O8^{+}$-plane, which T-dualizes to a pair of $O7^{+}$-planes in type IIB string theory. Such orientifolds lead to the phenomena of ``frozen singularities'' in F-theory \cite{Witten:1997bs, deBoer:2001px, Tachikawa:2015wka}. These are models in which the monodromy of the axio-dilaton around the brane is consistent with that of an appropriate $I_{n}^{\ast}$ singularity, but in which the corresponding gauge
algebra is \textit{not} $\mathfrak{so}(2n + 8)$.

Another not entirely unrelated phenomenon we have encountered in the construction of the $\mathcal{N} = (1,1)$ LSTs, as well as in some of the low rank LSTs are models in which the normal bundle of a curve on the base is torsion of finite order. To produce a Weierstrass model, we have found it necessary to impose specific restrictions on the order of these torsion bundles, though the M-theory realization of these A-type $\mathcal{N}=(1,1)$ theories suggests a whole family of models parameterized by rational theta angles \cite{Witten:1997kz}.

In fact, it is relatively straightforward to engineer all of the A-type $\mathcal{N}=(1,1)$ LSTs in type IIB,
and to lift this back to F-theory. For example, in type IIB language, we have a stack of $N$ D7-branes wrapped on a $T^{2}$. Switching on a
background value for a flat RR and NS two-form potential, we get additional theories parameterized by the ratio of these two periods. In F-theory language, we see this by a choice of how we resolve the affine node of the $\widehat{A}_{N+1}$ Dynkin diagram. In physical terms, this resolution comes from compactifying an 8D model on an additional $S^{1}$. Going down on a $T^2$, we have Wilson lines for this affine node along the A- and B-cycles of the $T^2$.

The presence of such background B-fields also suggests that similar effects
from discrete group actions may
also make an appearance in the construction of the $4 || 2$ type configurations.
For example, at the level of the effective field theory, we can consider a base:
\begin{equation}
2\overset{2}{2}2...2\label{Hitchorig}%
\end{equation}
with $I_N$-type fiber decorations on each $-2$ curve. Now, this field theory admits a $\mathbb{Z}_2$ automorphism in which we combine
the outer automorphism of the D-type base with an outer automorphism on the $\mathfrak{su}_{N}$ factors on the leftmost $-2$ curve
and the top $-2$ curve. At the level of gauge theory, the outer automorphisms of $\mathfrak{su}_{N}$ can take us to either an $\mathfrak{so}$ or $\mathfrak{sp}$-type algebra. Combining these two operations, we see that we get an effective field theory where there is one less tensor multiplet, and in which the BPS charge has doubled, and in which the Dirac pairing between this $\mathbb{Z}_2$ invariant combination and its neighbor is $2$, leaving us with a configuration of tensor multiplets $4 || 2 ... 2$. Taking into account the algebra assignment (i.e. the would-be fibers), we can in principle have either $\mathfrak{so}$ or $\mathfrak{sp}$ -type algebras, of which only the former is compatible with anomaly cancellation.\footnote{Let us recall that in F-theory, conjugation by an outer automorphism apparently leads to a geometrically ambiguous assignment for the quotient algebra. This ambiguity has been resolved by appealing to anomaly cancellation considerations, in which the opposite conclusion is reached, i.e. a non-split $I_{N}$ fiber realizes an $\mathfrak{sp}$-type algebra \cite{Aspinwall:2000kf}. However, effective field theory considerations suggest that when combined with a quotient on the tensor multiplets, there may be a generalization of this construction available in which we instead reach an $\mathfrak{so}$-type algebra. Incorporation of an $O7^{+}$ plane or of discrete B-fields might provide a route to such a generalization.} Similar considerations also apply for the LST tensor multiplet configurations such as $ 4  || 2 ... 2 1$.

Let us stress that effective field theory considerations do not directly inform us of the actual non-geometric realization of these models
in F-theory. Indeed, what is particularly remarkable is that \textit{even if} we allow these additional structures, the total number of additional candidate SCFTs and LSTs is quite small, with the vast majority being covered by geometric phases of F-theory. This suggests that whatever the mild deformation of known F-theory backgrounds are that produce these models, the structures encountered in this paper and in earlier work remain quite robust.

\section{Conclusions \label{sec:CONC}}

In this paper we have given a systematic approach to realizing supersymmetric little string theories
via compactifications of F-theory. Much as in the case of 6D SCFTs, these theories arise from working with
F-theory on a non-compact base, in which some collection of curves are simultaneously collapsed to small size.
The key difference with a 6D SCFT is that the intersection pairing for these curves defines a negative semidefinite
quadratic form on the lattice of string charges. So, in contrast to the case of SCFTs
the associated theories contain a dimensionful parameter which is naturally promoted to a non-dynamical tensor multiplet.
After spelling out all necessary conditions
to geometrically realize LSTs in F-theory, we have given a classification of all such theories. On the one hand, these theories can all be viewed
as arising from extending 6D SCFTs by one or more additional curves. As such, they also admit an atomic classification, much as in reference \cite{Heckman:2015bfa}. We have also seen that the general expectation that all 6D SCFTs embed in an LST is indeed realized via
the explicit embedding in an F-theory compactification. Finally, we have seen that
T-duality of an LST is realized via a double elliptic fibration in the corresponding
F-theory model. In the remainder of this section we discuss some avenues for future investigation.

Perhaps the most important issue left open by our analysis is the small gap between theories realized by geometric
phases of F-theory, and the list of effective field theories which can potentially complete to an LST (or SCFT). It
would be interesting to establish to what extent non-geometric deformations can enter in such F-theory models, and conversely,
how many of these putatively consistent LSTs (and SCFTs) are actually excluded by further non-trivial consistency conditions.

One of the key simplifications in our analysis of LSTs is that decoupling any curve in the base takes us back to a
collection of (possibly decoupled) 6D SCFTs and scale invariant theories (i.e. when we have free vector multiplets). This
strongly suggests that the common notions of renormalization group flows for local quantum field theories extend to \textit{non-local}
LSTs. Developing the details of such a structure would provide a rather striking vantage point on what it means to ``UV complete'' a
quantum field theory in the first place.

As a preliminary step in this direction, it is also natural to ask whether there is a notion of monotonic loss in the degrees of freedom in such conjectural flows from LSTs to SCFTs. For example, in many of the cases studied in this paper, we can weakly gauge both diffeomorphisms
as well as an $SU(2)$ field strength, which in the context of a 6D SCFT would be identified with the R-symmetry of the theory.
It is tempting to conjecture that there is a formal extension of conformal anomalies to these cases as well. It would be interesting
to study whether an extension of the methods presented in references \cite{Heckman:2015ola, Heckman:2015axa} (see also \cite{Cordova:2015fha})
would provide insight into such generalizations of renormalization group flows.

Finally, one of the hallmarks of the systems we have encountered is the appearance of an effective T-duality upon compactification on a further circle. Given that there are now concrete methods for extracting the partition functions for some 6D SCFTs (see e.g. \cite{Haghighat:2014pva}), it would be quite natural to study this and related structures for LSTs.

\newpage

\section*{Acknowledgements}

We thank D.S. Park, Y. Tachikawa and A. Tomasiello for helpful discussions.
LB thanks IISc and ICTS, Bangalore for hospitality during the Strings 2015 conference where a part of this work was completed.
MDZ, JJH, TR and CV thank the Simons Center for Geometry
and Physics 2015 summer workshop for hospitality during part of this work.
JJH also thanks the theory groups at Columbia
University, the ITS at the CUNY\ graduate center, and the CCPP at NYU for
hospitality during the completion of this work.
The work of LB is supported by the Perimeter Institute for Theoretical Physics. Research at Perimeter Institute is supported by the Government of Canada through Industry Canada and by the Province of Ontario through the Ministry of Economic Development and Innovation.
The work of MDZ, TR and CV is supported by NSF grant PHY-1067976.
TR is also supported by the NSF GRF under DGE-1144152.
The work of JJH is supported by NSF CAREER grant
PHY-1452037. JJH also acknowledges support from the Bahnson Fund at UNC Chapel Hill.
The work of DRM is supported by NSF grant PHY-1307513.


\appendix

\section{Brief Review of Anomaly Cancellation in F-theory}\label{anomalyagainandagain}

The allowed gauge algebras and matter content for a given tensor branch structure is heavily constrained by anomaly cancellation.  In 6D, anomalies are related to four-point amplitudes of external currents $J_a$ associated with continuous symmetry group $G_a$.  In such a four-point amplitude, insertions of external  currents $J_a$ for a given gauge group $G_a$ must come in pairs, so we need only consider the anomalies related to the four point functions $\langle J_a J_a J_a J_a \rangle $ (gauge anomaly cancellation) and $\langle J_a J_a J_b J_b \rangle $ (mixed gauge anomaly cancellation).

For a representation $R$ of some gauge group $G$, we introduce constants $Ind_R$, $x_R$, and $y_R$ relating the quadratic and quartic Casimirs of $G$ as follows:
\begin{equation}
\text{Tr}_RF^2= Ind_R \text{tr}F^2\,,~~~~\text{Tr}_RF^4= x_R \text{tr}F^4 + y_R  (\text{tr}F^2)^2. \label{def}
\end{equation}
Here, tr indicates the trace in a defining representation of the group\footnote{For $SU(N)$ and $Sp(N)$, the defining representation is simply the fundamental representation.  For $SO(5)$ and $SO(6)$, it is the spinor representation.  For $SO(N), N\geq 7$, it is the fundamental representation, though normalized with an additional factor of 2 so that Tr$_fF^2 = 2$tr$F^2$, Tr$_fF^4 = 2$tr$F^4$.  A complete list including exceptional gauge groups can be found in Table 2 of \cite{Grassi:2011hq}.}.

The constraints on gauge and mixed anomalies take the form \cite{Sadov:1996zm,Bershadsky:1997sb,Grassi:2000we,Grassi:2011hq}
\begin{align}
Ind_{Adj_a} - \displaystyle\sum_R Ind_{R_a} n_{R_a} &= 6 (10-n) \frac{\Omega_{IJ} a^I b_a^J}{\Omega_{IJ} a^I a^J}\\
y_{Adj_a} - \displaystyle\sum_R y_{R_a} n_{R_a} &= -3 (10-n) \frac{\Omega_{IJ} b_a^I b_a^J}{\Omega_{IJ} a^I a^J}\\
x_{Adj_a} - \displaystyle\sum_R x_{R_a} n_{R_a} &=0 \\
\displaystyle\sum_{R,R'} Ind_{R_a} Ind_{R_b'} n_{R_a R_b'} &= (10-n) \frac{\Omega_{IJ} b_a^I b_b^J}{ \Omega_{IJ} a^I a^J }.
\end{align}
Here, $\Omega_{IJ}$ is the natural metric on the space of antisymmetric tensors, and $a^I$, $b_a^J$ are related to the anomaly 8-form $I_8$ via
\begin{equation}
I_8 = \frac{1}{2} \Omega_{IJ} X^I X^J\,,~~~~X^I = \frac{1}{2} a^I \text{tr} R^2 + 2 b_a^I \text{tr} F_a^2.
\end{equation}
These field theoretic conditions can be translated into F-theory language as restrictions on the allowed gauge algebras and matter for a given base.  Any gauge algebra summand $\mf{g}_a$ in the theory is paired with a tensor multiplet, which is in turn associated with a curve $\Sigma_a$ in the base.  In these terms, the anomaly cancellation conditions become
\begin{align}
Ind_{Adj_a} - \displaystyle\sum_R Ind_{R_a} n_{R_a} = 6 (K \cdot \Sigma_a) &= 6(2g_a - 2 - \Sigma_a \cdot \Sigma_a)\label{firstanomaly}\\
y_{Adj_a} - \displaystyle\sum_R y_{R_a} n_{R_a} &= -3 (\Sigma_a \cdot \Sigma_a) \label{secondanomaly} \\
x_{Adj_a} - \displaystyle\sum_R x_{R_a} n_{R_a} &=0  \label{thirdanomaly} \\
\displaystyle\sum_{R,R'} Ind_{R_a} Ind_{R_b'} n_{R_a R_b'} &= \Sigma_a \cdot \Sigma_b. \label{fourthanomaly}
\end{align}
Here, $K$ is the canonical divisor of the base and $g_a$ is the genus of $\Sigma_a$.  The adjunction formula $K \cdot \Sigma_a = 2(g_a-1)- \Sigma_a \cdot \Sigma_a$ has been used in the second equality of (\ref{firstanomaly}).

Gauge anomaly cancellation may be used to constrain the gauge groups paired with a given tensor node, and mixed anomaly cancellation constrains the gauge groups allowed on neighboring tensor nodes.  However, tensors need not be paired with gauge groups.  In particular, curves of self-intersection $0$, $-1$, or $-2$ can be devoid of a gauge group entirely.  Curves of self-intersection $0$ cannot touch any other curves, so this case is relatively uninteresting.  Curves of self-intersection $-1$ and $-2$, on the other hand, can touch other curves.

\section{Matter for Singular Bases}\label{Appendix:Singularity}

In this Appendix, we review the relationship between matter and singularities.  Further information can be found in \cite{ Morrison:2011mb}.

Consider some curve $\Sigma$.  When this curve becomes singular, there are two distinct notions of genus.  Geometric genus, denoted $p_g$, is the topological genus of the curve after all singularities have been resolved.  Arithmetic genus, denoted $g$, is the quantity related to the intersection theory of the curve by
\begin{equation}
2g-2 = K\cdot \Sigma + \Sigma^2 .
\end{equation}
The arithmetic genus is the one that shows up in the adjunction formula and hence enters the anomaly cancellation equation (\ref{firstanomaly}).  These two notions of the genus are related via
\begin{equation}
g = p_g + \sum_P \frac{m_P(m_P-1)}{2}.
\end{equation}
Here, the sum runs over all singular points $P$ of the curve and $m_P$ is the multiplicity of the singularity at $P$.  For a curve with a nodal singularity (of Kodaira type $I_1$) or a cusp singularity (of Kodaira type $II$), there is a single singular point of multiplicity $m_P=2$.  Hence, in each of these cases we have $g = p_g +1$.  The only values of $g$ and $p_g$ that can actually arise in our classification are $g=1$ and $p_g=0$.

Although $p_g$ does not show up directly in the anomaly cancellation conditions, it still determines the F-theory matter content.  Namely, $p_g$ is precisely the number of adjoint hypermultiplets charged under the gauge algebra paired with this curve.  For \emph{smooth} LST bases, we see that there is one adjoint whenever the genus $g=p_g$ of the curve is 1.  For singular curves, on the other hand, the arithmetic genus $g$ can be 1 even without any adjoint hypermultiplets.  For curves with gauge algebra $\mf{su}(N)$, one instead finds symmetric and antisymmetric representations, as noted in \cite{Sadov:1996zm}.

\section{Novel DE Type Bases}\label{appx:DEbases}

Appendices B and C \cite{Heckman:2015bfa} provided a complete list of DE type bases that can be used to construct 6D SCFTs.  All of these bases can show up in LSTs as well when certain non-DE type links are suitably attached.  However, there are also some novel DE type bases, which blow down to 0.

We use an abbreviated notation to describe these bases.  Namely, we specify curves of self-intersection $-4$, $-6$, $-8$, and $-12$ according to
\begin{gather}
D \simeq 4 \\
E_6 \simeq 6 \\
E_7 \simeq 8 \\
E_8 \simeq 12
\end{gather}

The allowed types of conformal matter between DE type nodes are represented as follows:
\begin{gather}
D1D\simeq D\overset{1,1}{\oplus}D\label{DDpair}\\
E_{6}131E_{6}\simeq E_{6}\overset{2,2}{\oplus}E_{6}\\
E_{7}12321E_{6}\simeq E_{7}\overset{3,3}{\oplus}E_{6}\\
E_{7}12321E_{7}\simeq E_{7}\overset{3,3}{\oplus}E_{7}\\
E_{7}1231D\simeq E_{7}\overset{3,2}{\oplus}D\\
E_{8}12231D\simeq E_{8}\overset{4,2}{\oplus}D\\
E_{6}1315131E_{6}\simeq E_{6}\overset{3,3}{\bigcirc}E_{6}\\
E_{6}13151321E_{7}\simeq E_{6}\overset{3,4}{\oplus}E_{7}\\
E_{6}131513221E_{8}\simeq E_{6}\overset{3,5}{\oplus}E_{8}\\
E_{7}123151321E_{7}\simeq E_{7}\overset{4,4}{\oplus}E_{7}\\
E_{7}1231513221E_{8}\simeq E_{7}\overset{4,5}{\oplus}E_{8}\\
E_{8}12231513221E_{8}\simeq E_{8}\overset{5,5}{\oplus}E_{8}. \label{E8E8pair}%
\end{gather}

We omit the subscripts above the $\oplus$ when dealing with the minimal type of conformal matter:
\begin{equation}
D\oplus D\simeq D\overset{1,1}{\oplus}D
\end{equation}%
\begin{equation}
D\oplus E_{6}\simeq D\overset{2,2}{\oplus}E_{6}
\end{equation}%
\begin{equation}
D\oplus E_{7}\simeq D\overset{2,3}{\oplus}E_{7}
\end{equation}%
\begin{equation}
D\oplus E_{8}\simeq D\overset{2,4}{\oplus}E_{8}
\end{equation}%
\begin{equation}
E_{6}\oplus E_{6}\simeq E_{6}\overset{2,2}{\oplus}E_{6}
\end{equation}%
\begin{equation}
E_{6}\oplus E_{7}\simeq E_{6}\overset{3,3}{\oplus}E_{7}
\end{equation}%
\begin{equation}
E_{6}\oplus E_{8}\simeq E_{6}\overset{3,5}{\oplus}E_{8}
\end{equation}%
\begin{equation}
E_{7}\oplus E_{7}\simeq E_{7}\overset{3,3}{\oplus}E_{7}
\end{equation}%
\begin{equation}
E_{7}\oplus E_{8}\simeq E_{7}\overset{4,5}{\oplus}E_{8}
\end{equation}%
\begin{equation}
E_{8}\oplus E_{8}\simeq E_{8}\overset{5,5}{\oplus}E_{8}
\end{equation}

Using this notation, we now list the novel DE type bases for LSTs.  All of these bases are positive semi-definite with a single zero eigenvalue.  We begin with configurations with only $D$ nodes:
\begin{align}
&  D\overset{3,3}\oplus D \\
&  D\overset{3,3}\bigcirc D
\end{align}%

The configurations with only $E_6$ nodes:
\begin{align}
&  E_6\overset{5,5}\oplus E_6
\end{align}%

The configurations with $D$ and $E_6$:
 \begin{align}
&  D\overset{3,5}\oplus E_6 \\
& D \oplus E_6^{\oplus 2} \overset{3,3}\oplus E_6 \\
& D \oplus E_6^{\oplus 2} \overset{3,3}\bigcirc E_6 \\
& D \oplus E_6 \overset{3,4}\oplus E_6 \\
& D \oplus E_6 \overset{3,2}\oplus D \\
& D^{\oplus 2} \oplus E_6 \oplus D
\end{align}%

The configurations with $D$ and $E_7$:
 \begin{align}
&  D \overset{3,3}\oplus E_7^{\oplus n} \overset{3,3}\oplus D\,,~~~~n=1,2,... \\
&  D \oplus E_7^{\oplus 3} \overset{4,4}\oplus E_7 \\
&  D \oplus E_7^{\oplus 2} \overset{4,5}\oplus E_7 \\
& D^{\oplus 2} \oplus E_7 \oplus D \\
& D \overset{2,4}\oplus E_7 \oplus D \\
& D \oplus 1\overset{2}321 \oplus E_7
\end{align}%

The configurations with $D$ and $E_8$:
 \begin{align}
&  D\overset{3,5}\oplus E_8^{\oplus n}  \overset{5,3}\oplus D\,,~~~~n=1,2,... \\
& D^{\oplus 3} \oplus E_8 \oplus D^{\oplus 2} \\
& D^{\oplus 4} \oplus E_8 \oplus D
\end{align}%

The configurations with $E_6$ and $E_8$:
 \begin{align}
&  E_6 \overset{5,5}\oplus E_8^{\oplus n} \overset{5,5}\oplus E_6\,,~~~~n=1,2,... \\
&  E_6 \overset{4,5}\oplus E_8^{\oplus 2} \oplus E_6^{\oplus 2}
\end{align}%

The configurations with $D$, $E_6$ and $E_7$:
 \begin{align}
&  D \oplus E_7 \overset{4,4}\oplus E_6 \\
&  D \oplus E_7^{\oplus 2} \overset{4,3}\oplus E_6 \\
& D \oplus E_6^{\oplus 2} \overset{3,5}\oplus E_7 \\
& D \oplus E_6^{\oplus 3} \overset{3,4}\oplus E_7 \\
& D \oplus E_6^{\oplus 4} \oplus E_7 \\
& D \oplus  E_7^{\oplus 2} \oplus  E_6^{\oplus 2} \\
& D \oplus  E_6 \oplus E_7 \oplus  E_6^{\oplus 2}
\end{align}%

The configurations with $D$, $E_6$ and $E_8$:
 \begin{align}
&  D \overset{4,5}\oplus E_8^{\oplus n} \overset{5,5}\oplus E_6\,,~~~~n=1,2,... \\
& D \oplus E_6^{\oplus 6} \oplus E_8 \\
& D^{\oplus 2} \oplus E_8^{\oplus 2} \oplus E_6^{\oplus 2} \\
&D^{\oplus 3} \oplus E_8  \overset{5,4}\oplus E_6
\end{align}%

The configurations with $D$, $E_7$ and $E_8$:
 \begin{align}
&  D \oplus E_7^{\oplus 6} \overset{4,5}\oplus E_8  \\
& D^{\oplus 2} \oplus  E_8^{\oplus 3} \oplus  E_7^{\oplus 2}
\end{align}%

The configurations with $E_6$, $E_7$ and $E_8$:
 \begin{align}
&  E_6 \overset{4,5}\oplus E_8 \oplus E_7 \oplus E_6 \\
&  E_6 \overset{4,5}\oplus E_8^{\oplus 3} \oplus E_7^{\oplus 2}
\end{align}%

The configurations with $D$, $E_6$, $E_7$ and $E_8$:
 \begin{align}
& D \oplus E_8 \oplus E_7^{\oplus 2} \oplus E_6 \\
& D^{\oplus 2} \oplus E_8 \oplus E_7  \oplus E_6
\end{align}%

\newpage

\section{Novel Non-DE Type Bases}\label{appx:nonDEbases}

The following is the list of novel bases constructed solely from curves of self-intersection $-1$, $-2$, $-3$, and $-5$:

\begin{gather}
12....21 \\
5 \oplus 1^{\oplus 5}  \\
1\underset{1}{\overset{1}5}12 \\
21{\overset{1}5}12 \\
1312 \\
1\ovo31 \\
131\ovo512 \\
131 \underset{1}{\ovo5}12 \\
1231\ovo512 \\
1231 \underset{1}{\ovo5}12 \\
12231\ovo512 \\
12231 \underset{1}{\ovo5}12 \\
2\ovo31512 \\
2\ovo31\ovo51
\end{gather}
\begin{gather}
3131512 \\
3131\ovo51 \\
215131512 \\
215131\ovo51 \\
1\ovo5131\ovo51 \\
1\ovo5131512 \\
1\ovo5131\ovo51 \\
1315131512 \\
1315131\ovo51 \\
12315131512 \\
12315131\ovo51 \\
122315131512 \\
122315131\ovo51 \\
151231512 \\
151231\ovo51 \\
512231512 \\
512231\ovo51 \\
12321512 \\
12321\ovo51 \\
1321512 \\
1321\ovo51 \\
13\ovo22 \\
2\ov{2}2...21 \\
151\ovo322 \\
1512\ovo32 \\
512222 \\
151222 \\
3122 \\
1315122
\end{gather}
\begin{gather}
12315122 \\
122315122 \\
215122 \\
1\ovo5122 \\
13151\ovo32 \\
123151\ovo32 \\
1223151\ovo32 \\
2151\ovo32 \\
1\ovo51\ovo32 \\
3 2 1 3 2\\
1 2 2 3 1 3 2 1\\
2 3 1 2 3\\
1 2 3 1 3 2 2 1\\
1 3 1 3 2 2 1\\
1 5 1 2 2 3 1\\
1 3 2 1 3\\
1 2 3 1 3 1\\
1 3 1 3 1\\
3 1 2 3 2 1\\
3 1 3 2 1 5 1\\
2 3 1 3 1 5 1\\
1 5 1 2 3 2 1 5 1\\
1 2 3 2 1 5 1 3 2 1\\
5 1 2 3 1 3 2\\
3 2 1 5 1 2 3\\
1 2 3 1 5 1 2 3 1\\
1 3 1 5 1 2 3 1\\
1 3 2 1 5 1 3 2 2 1\\
5 1 3 1 3 2 2\\
1 3 2 1 5 1 3 1\\
3 2 2 1 5 1 3
\end{gather}
\begin{gather}
1 5 1 2 3 1 3\\
5 1 2 2 3 1 3\\
1 2 3 1 5 1 2 3 2 1\\
1 2 2 3 1 5 1 2 3 2 1\\
1 3 1 5 1 2 3 2 1\\
1 2 3 1 5 1 3 2 1 5 1\\
1 3 1 5 1 3 2 1 5 1\\
3 1 5 1 2 3 1 5 1\\
1 2 2 3 1 5 1 3 2 2 1 5\\
3 2 1 5 1 3 1 5 1\\
3 1 5 1 2 3 2 1 5\\
5 1 3 2 1 5 1 3 2\\
5 1 2 2 3 1 5 1 3 2 1\\
5 1 2 3 1 5 1 2 3\\
1 5 1 2 3 1 5 1 3 2 2 1\\
3 1 3 1 5 1 3 2 2 1\\
1 5 1 2 3 1 5 1 3 1\\
5 1 2 2 3 1 5 1 3 1\\
5 1 2 3 2 1 5 1 3\\
1 2 3 1 5 1 3 1 3\\
1 3 1 5 1 3 1 3\\
5 1 3 1 5 1 2 3 2\\
1 5 1 3 1 5 1 3 2 1 5\\
1 5 1 3 1 5 1 3 1 5 1\\
1 2 2 3 1 5 1 3 1 5 1 3 2 1\\
1 2 3 1 5 1 3 1 5 1 3 2 2 1\\
1 3 1 5 1 3 1 5 1 3 2 2 1\\
1 2 3 1 5 1 3 1 5 1 3 1\\
1 3 1 5 1 3 1 5 1 3 1\\
5 1 3 1 5 1 3 1 5 1 3
\end{gather}

\section{Novel Gluings} \label{appx:NOVEL}
At times, non-DE type side links or noble atoms can attach to DE type nodes in ways they could not for 6D SCFTs.  For instance, the side link 2151321 could never attach to a $D$ node in a 6D SCFT, since it induces four blow-downs on the $-4$ curve.  However, this is allowed for LSTs.  The full list of these novel gluings of one side link to a single node is as follows:

\bigskip

For gluing to a D-type node (i.e. a $-4$ curve), we have:
\begin{gather}
2221 \oplus D\\
2151321 \oplus D \\
3215131 \oplus D \\
1512321 \oplus D \\
3151231 \oplus D \\
31\ovo5131 \oplus D \\
22\ovo31 \oplus D \\
1\ovo51321 \oplus D \\
151315131 \oplus D \\
512315131 \oplus D
\end{gather}
For gluing to an $E_6$-type node (i.e. a $-6$ curve), we have:
\begin{gather}
222221 \oplus E_6 \\
2231\ovo5131 \oplus E_6 \\
231512321 \oplus E_6 \\
232151321 \oplus E_6 \\
321513221 \oplus E_6 \\
31\ovo513221 \oplus E_6 \\
231\ovo51321 \oplus E_6 \\
15131513221 \oplus E_6 \\
51231513221 \oplus E_6 \\
3151315121 \oplus E_6 \\
23151315131 \oplus E_6
\end{gather}
For gluing to an $E_7$-type node (i.e. a $-8$ curve), we have:
\begin{gather}
\underset{8}{\underbrace{22222221}} \oplus E_7 \\
2231513151321 \oplus E_7 \\
2231\ovo513221 \oplus E_7 \\
2231\ovo513221 \oplus E_7
\end{gather}
while for an $E_8$-type node (i.e. a $-12$ curve), we have:
\begin{gather}
\underset{12}{\underbrace{222222222221}}\oplus E_8.
\end{gather}
We remark that in the case of the $E_7$ and $E_8$ nodes, we can also delete some of the aforementioned curves, using instead a ``primed'' node, (i.e. by adding a small instanton link elsewhere).

\section{T-Duality in the $1,2,\dots,2,1$ Model \label{appx:Texamp}}

Let us consider the LST whose F-theory base $B$ has a chain of rational curves
consisting of two curves of self-intersection
$-1$ at the ends of the chain, and $k\ge0$ curves of self-intersection $-2$
in between meet each other.  The union of those $k+2$ curves on the base
deforms to a nonsingular rational curve of self-intersection $0$, and
provides a fibration $\pi:B \to C$ on the base whose general fiber is
$\mathbb{P}^1$.
The total space of the elliptic fibration looks like a one parameter
family of elliptic K3 surfaces (over general fibers of $\pi$)
degenerating to a pair of $dP_9$'s with $k$
intermediate elliptic ruled surfaces (over $\pi^{-1}(0)$), which (when $k=0$) is precisely
the degeneration which appears in the analysis of heterotic / F-theory duality in
reference \cite{Morrison:1996pp}.

Remarkably, the total space of this elliptic fibration admits a {\em second}\/
elliptic fibration, at least birationally.  The computation which shows
this was written out in reference \cite{McOrist:2010jw}, although it has
some earlier antecedents in the math literature
\cite{k3Picard,MR1013073,math.AG/0602146,MR2427457}.\footnote{There was a subsequent
extension of this computation to a more general case
\cite{MR2427457,arXiv:1004.3503,K3-modular-heterotic}, which will undoubtedly
be useful for understanding additional T-dualities of LSTs.}

Consider a base of the form $\mathbb{P}^1\times \mathbb{C}$
where the homogeneous coordinates on $\mathbb{P}^1$ are $[\sigma,\tau]$
on the coordinate on $\mathbb{C}$ is $\psi$.  We write a Weierstrass equation
of the form
\begin{equation}
 Y^2 = X^3 + a \sigma^4\tau^4 X + (\psi^{k+1} \sigma^5\tau^7 + c \sigma^6\tau^6
+\sigma^7\tau^5) , \label{eq:E8}
\end{equation}
where $a$ and $c$ are constants.
(In \cite{McOrist:2010jw} only the $\tau=1$ affine chart appears.)  Notice that we get
a small instanton with instanton number $k+1$
when $\psi=\sigma=0$, so to get the $1,2,\dots,2,1$ model we should blowup
the point $\psi=\sigma=0$ $k+1$ times. We keep this implicit in what follows.

Note that we have Kodaira type $II^*$ at both $\sigma=0$ and $\tau=0$,
and since those represent non-compact curves, we see $E_8\times E_8$
global symmetry.

Now there is a remarkable coordinate change:
\begin{align}
X &= st^{-5}x^2 \\
Y &= t^{-8}x^2y \\
\sigma &= t^{-3}x \\
\tau &= t.
\end{align}
(This is only a ``rational map'' because of the division by powers of $t$.)
The result of the substitution is
\begin{equation}
t^{-16}x^4y^2 = s^3t^{-15}x^6 + a  s t^{-13} x^6 + (\psi^{k+1} t^{-8}x^5
+ c t^{-12}x^6 + t^{-16}x^7).
\end{equation}
If we multiply the resulting equation by $t^{16}/x^4$, we obtain
\begin{align}
y^2
&= s^3tx^2 + a  s t^3 x^2 + (\psi^{k+1} t^8x
+ c t^4x^2 + x^3)\\
&= x^3 + (s^3t+ast^3+ct^4)x^2 + \psi^{k+1} t^8x.
\end{align}
We interpret this as a family elliptic curves over
$\mathbb{P}^1_{[s,t]}\times \mathbb{C}$.  (Note that this is a very
different base, with coordinates $[s,t]$ which mix the former base
and fiber coordinates.)
Remarkably, this is the equation for the F-theory dual of
the $Spin(32) / \mathbb{Z}_2$ heterotic string with a small instanton at $t=\psi=0$
(as derived in \cite{Morrison:1996pp, Aspinwall:1997ye})!  We have a
global symmetry algebra $\mathfrak{so}(32)$ along $t=0$.
Note that we should also blow up $t=\psi=0$.

Thus, in a quite subtle way there is T-duality for this pair of models,
in which the two different elliptic fibrations on the semi-local total
space are exchanged, after a birational change. Note that
the coordinate change given above is (rationally) invertible.
The inverse is:
\begin{align}
x &= \sigma \tau^3 \\
y &= \sigma^{-2}\tau Y\\
s&= \sigma^{-2}\tau^{-1} X\\
t &= \tau.
\end{align}

\section{F-Theory Construction of $\mathcal{N} = (1,1)$ LSTs}
\label{app:F1-1}

The M-theory construction for little string theories with
$\mathcal{N}=(1,1)$ supersymmetry is described in \cite{Witten:1997kz}.
Such a theory can be seen as M-theory compactified on a spacetime
of the form $(\mathbb{C}^2\times S^1)/\Gamma$ for $\Gamma\subset SU(2)$
a finite subgroup.  The action on $S^1$ is by rotations, and there is
a subgroup $\Gamma'$ of $\Gamma$ which acts trivially on $S^1$, leading
to a short exact sequence of groups
\begin{equation}
0 \to \Gamma' \to \Gamma \to \mathbb{Z}_r \to 0 .
\end{equation}
Geometrically, there is an action of $\mathbb{Z}_r$ on the ALE space
$\mathbb{C}^2/\Gamma'$ corresponding to an automorphism of the corresponding
Lie group; the gauge group of the little string theory is the
subgroup commuting with the outer automorphism.

The mathematical description of these groups has been known for a long
time, and is nicely summarized in a table on p.~376 of \cite{Reid:ypg}
which we reproduce as Table~\ref{tab:reid}.  We have made a minor correction
to the table (already noted in footnote 15 of \cite{Witten:1997kz}),
and we have added the information about the gauge group and theta angle
as discussed in \cite{Witten:1997kz}.
In the table, the cyclic group action on $\mathbb{C}^3$ with coordinates
$(x,y,z)$ is described in terms of exponents $(\frac ar,\frac br, \frac cr)$
of the generators; the notation also indicates how the equation transforms
under the action.

\begin{table}[b]
\begin{center}
\begin{tabular}{llllllc}
&$r$&Type&$\varphi$&Description&Gauge Group&$\theta$\\ \hline
(1)&any&$\frac1r(1,-1,0;0)$&$xy+z^n$&
$A_{n-1} \overset{r\text{-to-}1}{\longrightarrow}A_{rn-1}$
&$SU(n)$& ${}\in\pi\mathbb{Q}$\\
(2)&$4$&$\frac14(1,3,2;2)$&$x^2+y^2+z^{2n-1}$&
$A_{2n} \overset{4\text{-to-}1}{\longrightarrow}D_{2n+3}$
&$Sp(n)$ & $\pi$ \\
(3)&$2$&$\frac12(0,1,1;0)$&$x^2+y^2+z^{2n}$&
$A_{2n-1} \overset{2\text{-to-}1}{\longrightarrow}D_{n+2}$
&$Sp(n)$ & $0$\\
(4)&$3$&$\frac13(0,1,2;0)$&$x^2+y^3+z^3$&
$D_{4} \overset{3\text{-to-}1}{\longrightarrow}E_{6}$
&$G_2$ & $0$\\
(5)&$2$&$\frac12(1,1,0;0)$&$x^2+y^2z+z^n$&
$D_{n+1} \overset{2\text{-to-}1}{\longrightarrow}D_{2n}$
&$SO(2n+1)$ & $0$\\
(6)&$2$&$\frac12(1,0,1;0)$&$x^2+y^3+z^4$&
$E_{6} \overset{2\text{-to-}1}{\longrightarrow}E_{7}$
&$F_4$ & $0$\\
\end{tabular}

\quad

\caption{Cyclic actions on ALE spaces}
\label{tab:reid}
\end{center}
\end{table}

To construct these theories using F-theory, we use a base $B$ which is
a neighborhood of an elliptic curve $\Sigma$ whose normal bundle
is a torsion line bundle of order $r$.  The base has a finite unramified
cover of degree $r$ which is a product $\Sigma\times \mathbb{C}$, and
the cyclic group $\mathbb{Z}_r$ will act on the elliptic fibration over
$\Sigma\times\mathbb{C}$ (which we take to be in Weierstrass form).
Thus, the classification is analogous -- we
must find cyclic group actions on Weierstrass elliptic fibrations
which induce the corresponding actions on the ADE singularities.

Note that, as observed in section \ref{sec:RANKZERO},
 this construction requires $r\in\{ 2, 3, 4,
6\}$, so that most of the instances of case (1) are ruled out.
In fact, we have been unable to find a conventional F-theory construction of
any instance of case (1) (which would correspond, in the interpretation
of \cite{Witten:1997kz}, to an $SU(n)$ theory with a rational theta angle).
Instead, as mentioned in section~\ref{sec:towards-embedding},
we anticipate an F-theory construction for these models
involving B-field expectation values.

In Table~\ref{tab:equation-action}, we present explicit forms of
these group actions in cases  (3), (4),
and (6), using a Weierstrass equation $\varphi$ with variables $(x,y,t)$.  The quotient
can be described in terms of a Weierstrass equation $\Phi$ whose variables
$(X,Y,T)$ are expressed in terms of $(x,y,t)$ in the table.

\begin{table}
\begin{center}
\begin{tabular}{lllll}
& Type & $\varphi$ & $(X,Y,T)$ & $\Phi$\\ \hline
(3) & $\frac12(0,1,1)$ & $-y^2+x^3+ux^2+wt^{2n}$ & $(t^2x,t^3y,t^2)$ & $-Y^2+X^3+uTX^2 + wT^{n+3}$ \\
(4) & $\frac13(1,0,2)$ &$-y^2+x^3+wt^3$ & $(t^4x,t^6y,t^3)$ & $-Y^2+X^3+wT^5$\\
(6) & $\frac12(0,1,1)$ &$-y^2+x^3+wt^4$ & $(t^2x,t^3y,t^2)$ & $-Y^2+X^3+wT^5$\\
\end{tabular}

\quad

\quad

\caption{Group actions on Weierstrass models.  Here $u$ and $w$ represent invariant functions
of $t$ which do not vanish at $t=0$.}
\label{tab:equation-action}
\end{center}
\end{table}

We now explain the geometry of the group actions by means of figures
illustrating cases (2), (3), (4), and (6).
We will explain the cases in the reverse  order from the
one given in Table~\ref{tab:reid}.

\begin{figure}[t!]
\begin{center}
\includegraphics{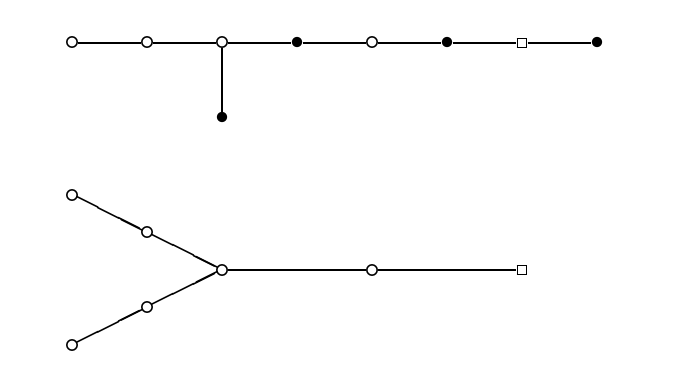}
\end{center}

\caption{$F_4$}
\label{figure-F4}
\end{figure}

We begin with case (6) leading to gauge symmetry $F_4$, illustrated in
Figure~\ref{figure-F4}.  The M-theory construction only involved the
Dynkin diagram of the singularities, but here we must consider the
entire Kodaira fiber as represented by the affine Dynkin diagram.  Thus,
in the bottom half of Figure~\ref{figure-F4} we see $\widehat{E}_6$, with
the affine node represented by a square rather than a circle.  This
diagram has been obtained as a double cover of the affine $E_8$
diagram
shown in the top half of Figure~\ref{figure-F4}, in which the solid
circles represent curves along which we branch as we construct a double
cover.  The two curves on the left are duplicated in the bottom half
because they don't meet the trivalent vertex at a branch point of the
double cover.  After taking the double cover, the solid curves have become
$-1$ curves and are to be blown down.  (Alternatively, the solid curves
can be contracted to $A_1$ singularities prior to taking the double cover.)

Note that the quotient involves an affine $E_8$ diagram rather than an
affine $E_7$ diagram.  This is because the $\mathbb{Z}_2$ action on
the ``extra'' curve in the Kodaira fiber (corresponding to the image of
the affine
node) has two fixed points, one giving
an $E_7$ singularity and the other giving an $A_1$ singularity.  The two
together fit into an affine $E_8$ diagram.

\begin{figure}[t!]
\begin{center}
\includegraphics{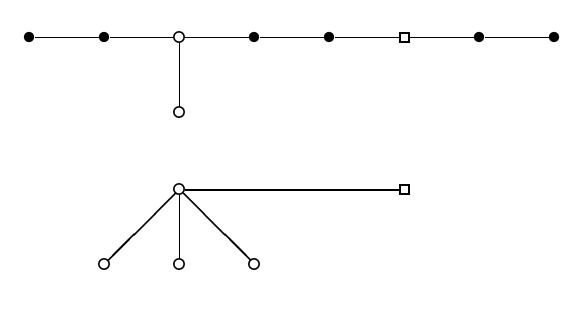}
\end{center}

\caption{$G_2$}
\label{figure-G2}
\end{figure}

We next treat case (4) leading to gauge symmetry $G_2$,
depicted in Figure~\ref{figure-G2}.  The bottom half of the figure
this time is $\widehat{D}_4$ with the affine node again represented by
a square.
This
diagram has been obtained as a {\em triple}\/ cover of the affine $E_8$
diagram
shown in the top half of Figure~\ref{figure-G2}, in which the solid
circles represent curves along which we branch as we construct a triple
cover.  (For an adjacent pair of solid curves, the branching is via $1/3$ on
one curve and $2/3$ on the other.)
The lowest curve in the top half of the diagram is triplicated in the
bottom half because it doesn't
meet the trivalent vertex at a branch point of the
triple cover.  After taking the triple cover, the solid curves have become
$-1$, $-2$ pairs  and are to be blown down.  (Alternatively, the solid curves
can be contracted to $A_2$ singularities prior to taking the triple cover.)
Note that the $\mathbb{Z}_3$ action on the image of the affine node
has two fixed points, one giving an $E_6$ singularity and the other
giving an $A_2$ singularity; the two together fit into an affine
$E_8$ diagram.

\begin{figure}[t!]
\begin{center}
\includegraphics{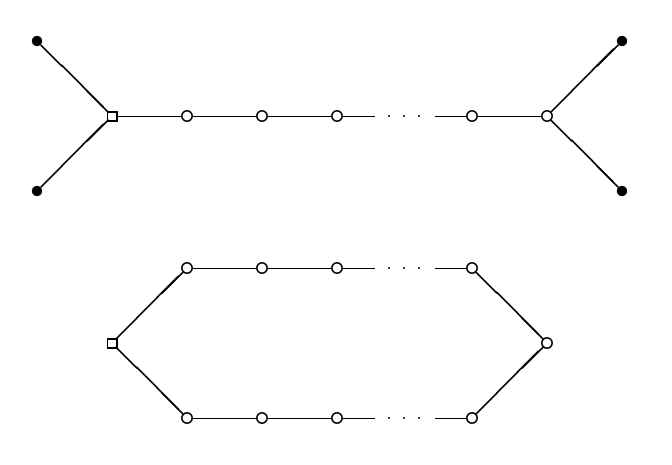}
\end{center}

\caption{$Sp(n)$, $\theta=0$}
\label{fig:Sp}
\end{figure}

Now we consider case (3), depicted in Figure~\ref{fig:Sp}.  The bottom half
is $\widehat{A}_{2n-1}$, obtained as a branched double cover of the
top half, which is $\widehat{D}_{n+4}$.  The square denotes the affine
node (or its image), and the solid circles in the top half denote
curves along which the double cover is branched.  This time, the
$\mathbb{Z}_2$ action on the the affine node has two fixed points
unrelated to the singularity we are studying, creating the two extra
solid curves at the left of the diagram.

\begin{figure}[t!]
\begin{center}
\includegraphics{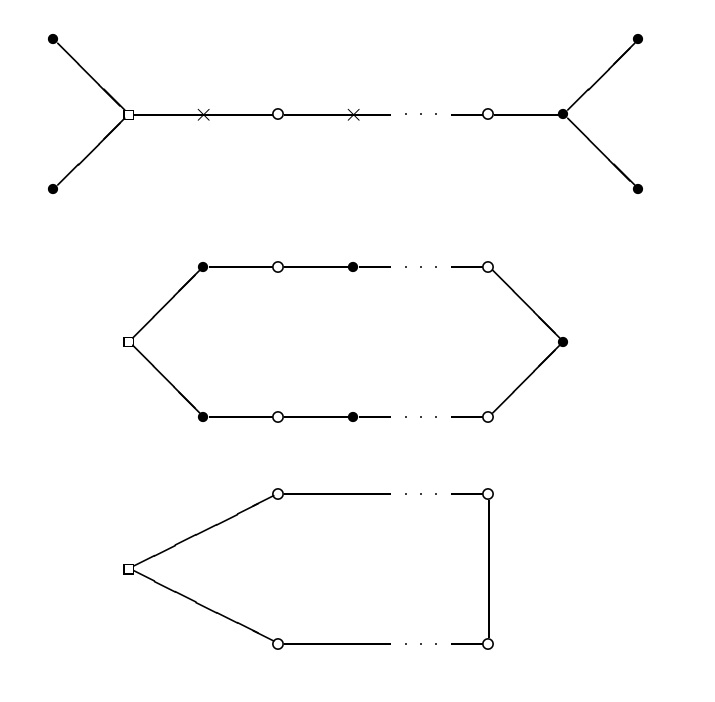}
\end{center}

\caption{$Sp(n)$, $\theta=\pi$}
\label{fig:Sp-theta}
\end{figure}

Finally, we treat case (2), depicted in Figure~\ref{fig:Sp-theta}, which is
the most complicated case.  Here, we have a $\mathbb{Z}_4$ action
on $\widehat{A}_{2n}$ at the bottom of the figure, and
we describe the quotient in two stages:  the quotient by the $\mathbb{Z}_2$
subgroup (shown in the middle of the figure as an $\widehat{A}_{4n+1}$
diagram) and the quotient by
the full $\mathbb{Z}_4$ (shown at the top of the figure as a
$\widehat{D}_{2n+5}$ diagram).  The diagram
at the top includes a resolved $A_3$ singularity (three connected solid dots),
two resolved $A_1$ singularities (at the far left)
which are to be branched along
during the first double
cover, and points marked by $\times$ which are $A_1$ singularities not
branched during the first double cover (from top to middle) but branched
during the second double cover (from middle to bottom).

\clearpage

\bibliographystyle{utphys}
\bibliography{LST}

\end{document}